\documentclass[lineno]{jfm}
\usepackage{graphicx}
\usepackage{mathtools}
\usepackage{newtxtext}
\usepackage{newtxmath}
\usepackage{natbib}
\usepackage{hyperref}
\usepackage{bm}
\usepackage{xcolor}
\usepackage{cases}
\usepackage{overpic}
\hypersetup{
    colorlinks = true,
    urlcolor   = blue,
    citecolor  = black,
}

\newcommand{\RomanNumeralCaps}[1]
\linenumbers

\newcommand\Wi{\mathit{Wi}}
\newcommand\Wicr{\mathit{Wi}_{\rm cr}}
\newcommand\Lscr{\mathscr{L}}
\newcommand\Req{R_{\mathrm{eq}}}
\newcommand\Qeq{Q_{\mathrm{eq}}}
\newcommand\Rmax{R_m}
\newcommand\rmax{r_m}
\newcommand\ratio{\gamma}

\title{Polymers in turbulence: stretching statistics and the role of extreme strain-rate fluctuations}
\author{Jason R. Picardo\aff{1}
  \corresp{\email{jrpicardo@che.iitb.ac.in}; {Also, Associate, International Centre for Theoretical Sciences, TIFR, India.}},
Emmanuel L. C. VI M. Plan\aff{2}
 \and Dario Vincenzi\aff{3}\corresp{{Also, Associate, International Centre for Theoretical Sciences, TIFR, India.}}}

\affiliation{\aff{1}Department of Chemical Engineering, Indian Institute of Technology Bombay, Mumbai, 400076, India
\aff{2}{Hanoi School of Business and Management, Vietnam National University, Ha Noi, 100 000, Vietnam}
\aff{3}Universit\'e C\^ote d'Azur, CNRS, LJAD, 06000 Nice, France
}

\begin{document}
\maketitle

\begin{abstract}
Polymers in a turbulent flow are stretched out by the fluctuating velocity gradient and exhibit a broad distribution of extensions $R$; the \textit{stationary} probability distribution function (p.d.f.) of $R$ has a power-law tail with an exponent that increases with the Weissenberg number $\Wi$, a nondimensional measure of polymer elasticity. This study addresses the following questions: (i) What is the role of the non-Gaussian statistics of the turbulent velocity gradient on polymer stretching? (ii) How does the p.d.f. of $R$ evolve to its asymptotic stationary form? Our analysis is based on simulations of the dynamics of finitely-extensible bead-spring dumbbells and chains, in the extremely dilute limit, that are transported in a homogeneous and isotropic turbulent flow, as well as in a Gaussian random flow. First, we recall the large deviations theory of polymer stretching, and illustrate its application. Then, we show that while the turbulent flow is more effective at stretching small-$\Wi$ stiff polymers, the Gaussian flow is more effective for high-$\Wi$ polymers. This suggests that high-$\Wi$ polymers (with large relaxation times) are primarily stretched by the cumulative effect of moderate strain-rate events, rather than by short-lived extreme-valued strain rates; we confirm this behaviour by analysing the persistence time of polymers in stretched states. Next, we show that, beginning from a distribution of coiled polymers, the p.d.f. of $R$ exhibits two distinct regimes of evolution. At low to moderate $\Wi$, the p.d.f. quickly develops a power-law tail with an exponent that evolves in time and approaches its stationary value exponentially. This result is supported by an asymptotic analysis of a stochastic model. At high $\Wi$,
the rapid stretching of polymers first produces a peak in the p.d.f. near their maximum extension; a power law with a constant exponent then emerges and expands its range towards smaller $R$. The time scales of equilibration, measured as a function of $\Wi$, point to a critical slowing down at the coil-stretch transition. Importantly, these results show no qualitative change when chains in a turbulent flow are replaced by dumbbells in a Gaussian flow, thereby supporting the use of the latter for reduced-order modelling.
\end{abstract}



\section{Introduction}
\label{sec:intro}

The non-Newtonian behaviour of turbulent polymer solutions results from
the stretching of polymer molecules dissolved in the fluid.
A detailed understanding of the statistics of polymer stretching in turbulent flows is, therefore, essential
to explain phenomena such as turbulent drag reduction \citep{g14,bc18,x19} and elastic turbulence \citep{steinberg21}. On the flip side, extreme stretching causes the mechanical scission of polymers, following which all non-Newtonian effects are lost \citep{s20}. To accurately model scission and the consequent loss of viscoelasticity, we must understand how the turbulent flow stretches out individual polymers to large extensions. 

The strain rate in a turbulent flow fluctuates in magnitude and orientation. In addition, vorticity constantly rotates the polymers and prevents them from persistently aligning with the stretching eigendirection of the strain-rate tensor.
Nonetheless, \cite{l73} predicted that a turbulent flow can stretch polymers, if the product of the mean-square strain rate and its Lagrangian integral time exceeds the inverse of the elastic relaxation time of the polymers.
The Lagrangian numerical simulations of \cite{mksh93} confirmed the
unravelling of individual bead-spring chains in a turbulent channel flow, while \cite{gsprl01} provided experimental evidence of this phenomenon in elastic turbulence: The transition from a laminar shear flow to a chaotic flow was found to coincide with a dramatic increase in the polymer-induced shear stress.

A systematic theory of polymer stretching in turbulent flows was developed by \cite*{bfl00,bfl01} and \cite{c00} using the methods of
dynamical systems.
One of the main results is that the end-to-end extension $R$ of the polymer has a stationary probability density function (p.d.f.), $p(R)$, that
behaves as a power law $R^{-1-\alpha}$ for $R$ between the equilibrium and contour lengths of the polymer.
The exponent $\alpha$ is a function of the Weissenberg number $\Wi$, defined as the product of the Lyapunov exponent
of the flow and the polymer relaxation time; this relationship is expressed in terms of the Cram\'er function
that describes the large deviations of the stretching rate of the flow.  
Note that the power-law behaviour of $p(R)$ is a distinctive feature of turbulent flows---the distribution 
of polymer extensions remains broad even at large strain rates---and is not observed
in laminar, extensional flows \citep*{psc97,nk99,s18}.

The power-law behavior of the p.d.f. of $R$ has been confirmed in various flow configurations:
experimentally by direct observation of individual polymers in elastic turbulence \citep*{gcs05,ls10,ls14};
and numerically in shear turbulence \citep*{eks02,pt05}, in two- and three-dimensional isotropic turbulence \citep*{bcm03,wg10,gpp15,rpm21},
and in turbulent channel and pipe flows \citep{bmpb12,sbgc22}.
This study further investigates the power-law behaviour of $p(R)$
by examining (i) how  polymer stretching is affected by the extreme velocity gradient fluctuations that characterize turbulence and (ii) how the p.d.f. of $R$ evolves from an initial distribution of coiled polymers to its stationary profile.

 Typically, the contour length of the polymer is smaller than the viscous scale of the turbulent flow, and so polymer deformation is entirely determined by the statistics of the velocity gradient (\citealt{idakcs02,sg03,za03};
\citealt*{gsk04}; \citealt{tdmsl04,ps07}). It is natural, therefore, to consider simplifying the simulation of polymer stretching by using a random velocity gradient model with appropriate statistics. This would facilitate the testing and development of advanced polymer models, beyond the simple dumbbell or freely-jointed chain, by avoiding expensive direct numerical simulations (DNS) of the carrier flow. However, one must first answer the question: Does stochastic modelling of the turbulent flow modify, in a fundamental way, the stretching dynamics of polymers?

In the broader context of the turbulent transport of anisotropic particles, several studies have shown that particle-orientation dynamics cannot be explained fully by using Gaussian models of the velocity gradient (see, \textit{e.g.},
\citealt{pw11,cm13}; \citealt*{gem14,ars20}).
Meanwhile, for single-polymer dynamics, Gaussian velocity-gradient models have undoubtedly been useful in gaining a qualitative understanding of stretching statistics \citep[\textit{e.g.}][]{sk92,ms97,c00,mav05,pt05,vwrp21}.
To our knowledge, however, the effect of the strongly non-Gaussian fluctuations of a turbulent velocity gradient
on the statistics of $R$, and in particular on the dependence of $\alpha$ upon $\Wi$, has not yet been 
investigated.
With this aim, we carry out Lagrangian numerical simulations of
finitely extensible nonlinear elastic (FENE) dumbbells
in three-dimensional
isotropic turbulence, as well as in a Gaussian model of the velocity gradient, and compare the statistics of $R$. This investigation has relevance beyond stochastic modelling, because flows without extreme strain-rates arise naturally in the context of polymer solutions---wherein polymer-feedback forces suppress extreme velocity-gradient fluctuations \citep*{prasad10,wg13,rll22}.

So far, most studies have focused on the stationary statistics of polymer extensions. 
However, the transient dynamics is important in situations where polymer scission 
occurs and the long-time stationary state may never be reached \citep{s20}.
In addition, the finite-time statistics of polymer stretching is relevant to
experimental measurements, which are necessarily limited in their duration, as well as to the calibration of numerical simulations.
The characteristic time required for polymers to equilibrate in a turbulent flow has been studied in 
stochastic models \citep*{mav05,cpv06} 
and  isotropic turbulence \citep{wg10}. A significant
slowing down of the stretching dynamics has been found near the coil--stretch transition.
This phenomenon is reminiscent of the slowing down observed in an extensional flow \citep{gs08}, but has a different origin. Specifically, it is not a consequence of the conformation hysteresis typical of extensional flows \citep{sbsc03,ssc04}, but rather arises from the strong heterogeneity of polymer configurations in the vicinity of the coil--stretch transition \citep{cpv06}.
Other than the equilibration time, little is known about how the p.d.f. of polymer extension approaches its 
asymptotic shape and,
in particular, whether or not the power-law behaviour
appears at earlier times. Here, this issue is studied by means of Lagrangian simulations of single-polymer dynamics in three-dimensional isotropic turbulence;
simulations for both a dumbbell and a bead-spring chain are compared. In addition, the numerical results are explained by using
a Fokker--Planck equation for the time-dependent p.d.f. of $R$.

The dumbbell model of polymers is presented in \S~\ref{sec:flows}, along with a description of the Lagrangian simulations that form the basis of this study. The random and turbulent carrier flows are also described there. Section~\ref{sec:stationary} focuses on the stationary p.d.f. of polymer extensions. In \S~\ref{sec:renewing}, the theory of
\cite{bfl00} and \cite{c00} is recalled briefly and is illustrated by using a renewing Couette flow. Section~\ref{sec:Gaussian}
then examines the effect of the non-Gaussian statistics of a turbulent velocity gradient on polymer stretching.
Section~\ref{sec:equilibration} is devoted to understanding the temporal evolution of the p.d.f. of polymer extensions, with the aid of a stochastic model. We then verify, in \S~\ref{sec:chain}, that the results obtained for the elastic dumbbell remain true even for chains with a larger number of beads.
Finally, we conclude in \S~\ref{sec:conclusions} with a summary of our study and a discussion of its implications.

\section{Lagrangian simulations in random and turbulent flows}\label{sec:flows}

\subsection{Dumbbell model}\label{sec:dumbbell}
We primarily model polymer molecules using the finitely extensible nonlinear elastic (FENE) dumbbell model \citep{bird,o96,g18}. With $\tau$ as the relaxation time of the polymer, $\Req$ as the equilibrium root-mean-square (r.m.s.) end-to-end length, and $\Rmax$ as the maximum length, the dynamics of the end-to-end separation vector $\bm R$ in $d$ dimensions
satisfies
\begin{equation}
\frac{\mathrm{d}\bm R}{\mathrm{d}t}=\bm\kappa(t)\cdot\bm R-f(R)\,\frac{\bm R}{2\tau}
+\sqrt{\frac{\Req^2}{\tau d}}\,\bm\xi(t),
\label{eq:dumbbell}
\end{equation}
where $\kappa_{ij}=\nabla_j u_i$ is the velocity gradient at the location of the centre of mass of the polymer,
$f(R)=\big(1-R^2/\Rmax^2\big)^{-1}$ defines the FENE spring force, and $\bm\xi(t)$ is $d$-dimensional white noise that accounts for thermal fluctuations. This noise would also make an appearance in the equation of motion for the centre of mass. However, its effect on the transport of the dumbbell is very small compared to the advection by the turbulent carrier flow and thus may be neglected. The motion of the centre of mass of the polymer is therefore treated like that of a tracer.

Given a Lagrangian time series of $\bm\kappa(t)$, \eqref{eq:dumbbell} is integrated using the Euler--Marujama method supplemented with the rejection algorithm 
proposed by \cite{o96}, which rejects those
time steps that yield extensions greater than $\Rmax\big(1-\sqrt{\mathrm{d}t/10}\big)^{1/2}$. Since the velocity gradient fluctuates, the numerical integration of 
\eqref{eq:dumbbell} does not present the same difficulties as in the case of a laminar extension
flow, and more sophisticated integration methods are not necessary. We have indeed checked that only a negligible fraction of time steps is rejected over a simulation. 

Throughout this paper, we study the dumbbell model with $\Req=1$ and $\Rmax=110$. The elastic relaxation time $\tau$ defines the non-dimensional Weissenberg number, $\Wi \equiv \lambda \tau$, where $ \lambda$ is the maximum Lyapunov exponent of the carrier flow. 
The Weissenberg number provides a non-dimensional measure of the elasticity of the polymer, with high-$Wi$ polymers being easily extensible. 
We consider a wide range of $\Wi$, from 0.3 to 40. 

Note that while the dumbbell model \eqref{eq:dumbbell} is used in most of this study, we do show in \S~\ref{sec:chain} that our results also hold true for bead-spring chains.

\subsection{Turbulent carrier flow}\label{sec:DNStraj}
 To study polymer stretching in a turbulent flow, we use a database of Lagrangian trajectories from a DNS of homogeneous isotropic incompressible turbulence (at Taylor-microscale Reynolds number $\mathit{Re}_\lambda\approx 111$), generated at ICTS, Bangalore \citep{jr17}. 
The DNS solves the incompressible Navier--Stokes equations, discretized on a periodic cube, using a standard fully-dealiased pseudo-spectral method with $512^3$ grid points. Time integration is performed using a second-order slaved Adams--Bashforth scheme. The motion of $9\times 10^5$ tracers is calculated using a second-order Runge--Kutta method for time integration; the fluid velocity at the location of a tracer is obtained from its value on the grid using trilinear interpolation. The velocity gradient $\bnabla\bm u$ is calculated along these trajectories and stored at intervals of $0.11 \tau_\eta$, where $\tau_\eta$ is the Kolmogorov time-scale (given below). 
This Lagrangian data provides the values of $\bm\kappa(t)$ for the integration of \eqref{eq:dumbbell} along $9\times 10^5$ trajectories, allowing us to obtain good statistics of single-polymer stretching dynamics.

The Lyapunov exponent of the flow, required for defining $\Wi$, is computed from these trajectories via the continuous $QR$ method, implemented using an Adams--Bashforth projected integrator 
along with the composite trapezoidal rule \citep[for further details see][]{drv97}.
We find $\lambda=0.136/\tau_\eta$
which is compatible with previous simulations of isotropic turbulence \citep{bbbcmt06}.
We also estimate the Lagrangian correlation time-scales of strain-rate and vorticity in the turbulent flow, $\tau_{S}$ and $\tau_\varOmega$, which serve as inputs to the Gaussian random model described in the next subsection. The rate-of-strain and rotation tensors are defined as
$\mathsfbi{S}=(\bnabla\bm u+\bnabla\bm u^\top)/2$ and
$\mathsfbi{\varOmega}=(\bnabla\bm u-\bnabla\bm u^\top)/2$, respectively.
The autocorrelation functions of $S_{11}$ and $\varOmega_{12}$ are calculated and found to display an approximately exponential
decay. Integrating these functions yields the integral time scales
$\tau_{S}=2.20\,\tau_\eta$ and $\tau_\varOmega=8.89\,\tau_\eta$,
in agreement with previous numerical simulations at comparable $R_\lambda$ \citep{y01}. 
The Kolmogorov time-scale $\tau_\eta$ is determined from $S_{11}$, using isotropy, as
$\tau_\eta=\left({15\langle S_{11}^2\rangle}\right)^{-1/2}=3.72\times 10^{-2}$.


\subsection{Gaussian random velocity gradient}
\label{sec:Gaussian-grad}
One of the goals of this study is to compare the stretching of polymers in a turbulent flow to that in a flow with Gaussian statistics, in order to determine the effect of extreme-valued fluctuations of the turbulent velocity gradient. For this, we use a Gaussian random velocity-gradient model to generate a time series of $\bm\kappa(t)$ for each polymer trajectory.
Following \citet*{bkl97}, we take $\bm\kappa(t)=\mathsfbi{S}(t)+\mathsfbi{\varOmega}(t)$ with
\begin{equation}
\mathsfbi{S}=\sqrt{3}\,A
\begin{pmatrix}
\frac{2\zeta_1}{\sqrt{3}} & \zeta_3 & \zeta_4
\\
\zeta_3 & -\frac{\zeta_1}{\sqrt{3}}+\zeta_2 & \zeta_5
\\
\zeta_4 & \zeta_5  & -\frac{\zeta_1}{\sqrt{3}}-\zeta_2
\end{pmatrix},
\qquad
\mathsfbi{\varOmega}=\sqrt{5}\,A
\begin{pmatrix}
0 & \varpi_1 & \varpi_2
\\
-\varpi_1 & 0 & \varpi_3
\\
-\varpi_2 & -\varpi_3  & 0
\end{pmatrix},
\end{equation}
where $A$ determines the magnitude of the velocity gradient and $\zeta_i(t)$ ($i=1,\dots,5$)
and $\varpi_i(t)$ ($i=1,2,3$)
are independent zero-mean unit-variance
Gaussian random variables with exponentially decaying autocorrelation 
functions and integral times $\tau_S$ and $\tau_\varOmega$, respectively.
Therefore, $S_{ij}$ and $\varOmega_{ij}$ are Gaussian variables such that
$\langle S_{ij}\rangle=\langle\varOmega_{ij}\rangle=0$,
\begin{equation}
\langle S_{ik}(t)S_{jl}(0)\rangle = 3A^2
\left(\delta_{ij}\delta_{kl}+\delta_{il}\delta_{jk}-\frac{2}{3}\delta_{ik}
\delta_{jl}\right) \, e^{-t/\tau_S},
\end{equation}
and
\begin{equation}
\langle \varOmega_{ik}(t)\varOmega_{jl}(0)\rangle = 
5A^2 \left(\delta_{ij}\delta_{kl}-\delta_{il}\delta_{jk}\right)
e^{-t/\tau_\varOmega}.
\end{equation}
As a consequence, $\langle\varOmega_{ij}\varOmega_{ij}\rangle=\langle S_{ij}S_{ij}\rangle$,
which reproduces the relation $\nu\langle\omega^2\rangle=\langle\epsilon\rangle$, where $\omega$ is the vorticity and $\epsilon$ is the energy dissipation rate \citep{f95}.
We take $\tau_S$ and $\tau_\varOmega$ to be the same as in the turbulent flow (\S~\ref{sec:DNStraj}). Furthermore, we set the coefficient 
$A= 2.538$ so as to obtain approximately the same Lyapunov exponent $\lambda$ as in the turbulent flow. As a consequence, the Kubo number 
$\mathit{Ku}=\lambda\tau_S$ is also nearly the same in the turbulent and Gaussian flows.

\section{Stationary statistics of polymer stretching}

We begin our study by considering the long-time, statistically stationary distribution of the end-to-end extension of polymers. We first recall the large deviations theory of \cite{bfl00} and \cite{c00}, which not only predicts a power-law tail in the p.d.f of $R$, but also provides a way to calculate the corresponding exponent for any chaotic carrier flow, given sufficient knowledge of its dynamical properties. We illustrate the predictive capability of the theory for a simple, analytically specified, renewing flow. Unfortunately, direct application of the theory to turbulent flows is impractical and one must typically resort to approximations, such as considering the turbulent flow to be decorrelated in time. Such considerations will naturally lead us to examine how and to what extent the time-correlated, non-Gaussian nature of turbulent flow statistics impacts polymer stretching.

\label{sec:stationary}

\subsection{Large deviations theory: illustration for a renewing flow}
\label{sec:renewing}

The theory of \cite{bfl00} and \cite{c00} is recalled here in terms of the generalized Lyapunov exponents,
rather than the Cram\'er function of the strain rate; the two properties are equivalent and related via a Legendre transform
\citep*[see][]{bcm03}.
If $\bm\ell(t)$ is a line element in a random flow, the Lyapunov exponent is defined as
\begin{equation}
\lambda = \lim_{t\to\infty}\frac{1}{t}\left\langle\ln\left[\frac{\ell(t)}{\ell(0)}\right]\right\rangle,
\end{equation}
where $\langle\cdot\rangle$ denotes the average over the statistics of the velocity field.
The $q$-th generalized Lyapunov exponent,
\begin{equation}
\label{eq:Lp}
\Lscr(q) = \lim_{t\to\infty}\frac{1}{t}\ln\left\langle\left[\frac{\ell(t)}{\ell(0)}\right]^q\right\rangle,
\end{equation}
gives the asymptotic exponential growth rate of the $q$-th moment of $\ell(t)$.
The function $\mathscr{L}(q) $ is positive and convex and satisfies  $\Lscr'(0)=\lambda$
\citep*[see, \textit{e.g.},][for more details]{cc2010}.
If the flow is incompressible,  then we also have $\Lscr(-d)=\Lscr(0)=0$, where $d$ is the space dimension \citep{zrms84}.

For the elastic dumbbell~\eqref{eq:dumbbell}, the p.d.f of the extension has a power-law form, $p(R) \sim R^{-1-\alpha}$ for $\Req \ll R\ll \Rmax$, with an exponent that satisfies
\begin{equation}
\label{eq:alpha}
\frac{\alpha}{2\Wi} = \frac{\Lscr(\alpha)}{\lambda}.
\end{equation}
Given the properties of $\Lscr(q)$, equation \eqref{eq:alpha} implies that $\alpha$ is a decreasing function of $\Wi$
that crosses zero at the critical value $\Wicr=1/2$ and saturates to $-d$ for very large $\Wi$. In the limit $\Rmax\to\infty$, the p.d.f. of $R$ ceases to be normalizable for $\Wi\geq \Wicr$, \textit{i.e.}  highly stretched configurations predominate, and so $\Wicr$ is taken to mark the coil--stretch transition.

In principle \eqref{eq:alpha} may be used to determine $\alpha$ as a function of $Wi$. However, in general, calculating the function $\Lscr(q)$ is very challenging. A useful approximation can be obtained in the vicinity of the coil--stretch transition by expanding about $q=0$:
$\Lscr(q)=\lambda q+\varDelta q^2/2+O(q^3)$
with $\varDelta = \int \left(\left\langle\zeta(t)\zeta(t')\right\rangle - \lambda^2\right)dt'$
and $\zeta(t)=\bm{R}\cdot\bm{\kappa}(t)\cdot\bm{R}/R^2$.
Substituting this quadratic expansion into \eqref{eq:alpha} yields
\begin{equation}
\label{eq:alpha-kraichnan}
\alpha=\frac{\lambda}{\varDelta}\left(\frac{1}{\Wi}-2\right)\quad \mathrm{for}\;Wi \to \Wicr.
\end{equation}
Interestingly, in the limiting case of a time-decorrelated flow, \eqref{eq:alpha-kraichnan} is accurate for all $\Wi$, since
$\Lscr(q)$ is quadratic for all $q$ \citep*{fgv01}.
Moreover,  in this case $\lambda/\Delta=d/2$, which further simplifies \eqref{eq:alpha-kraichnan} to yield:
\begin{equation}
\label{eq:alpha-kraichnan-d}
\alpha=\frac{d}{2}\left(\frac{1}{\Wi}-2\right).
\end{equation}
%
\begin{figure}
\centering
\begin{overpic}[width=.88\textwidth]{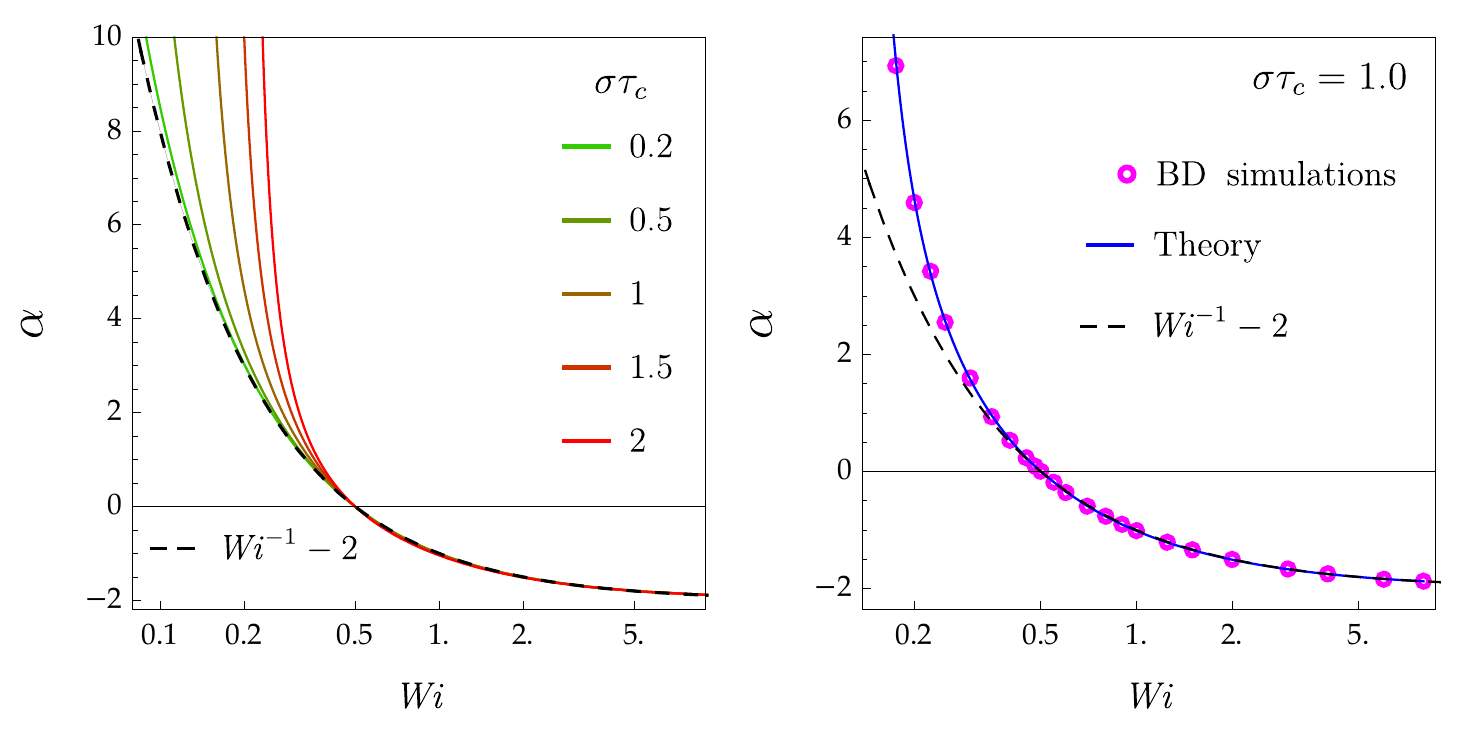}
\put (1,47) {(\textit{a})}
\put (51,47) {(\textit{b})}
\end{overpic}
\caption{(\textit{a}) Slope of $p(R)$ as a function of $\Wi$, as predicted by the large deviations theory, for the renewing Couette flow with various values of $\sigma \tau_c$. The dashed line is the prediction \eqref{eq:alpha-kraichnan-d} for a time-decorrelated flow ($\tau_c = 0$) with $d = 2$.
 (\textit{b}) Comparison of the large deviations theory (solid line) with BD simulations of the dumbbell model (markers), for the renewing Couette flow with $\sigma=10$ and $\tau_c=0.1$. The decorrelated flow (dashed line) is seen to provide a good approximation for $\Wi$ near to $\Wi_\mathrm{cr} = 1/2$ and beyond.
}
\label{fig:couette}
\end{figure}%
To obtain $\alpha$ for polymers in a general chaotic flow and for arbitrary $\Wi$, we must measure $\Lscr(q)$; due to statistical errors this is especially difficult for values of $q$ that are negative or large and positive \citep{v10}. Thus, past studies have been restricted to values of $\Wi$ sufficiently near $\Wicr$ so that $\alpha$ does not deviate far from zero \citep{gcs05,bmpb12}. In order to illustrate the validity
of \eqref{eq:alpha} over a wider range of $\Wi$, we now consider a renewing (or renovating) random flow, for which $\Lscr(q)$ can be calculated easily \citep{cg95}. To generate this flow, the time axis is divided into intervals $\mathscr{I}_n=[t_{n},t_{n+1})$ with $t_n=n\tau_c$ and
$n=1,2,\dots$. The velocity field changes randomly at the beginning of each interval and then remains frozen for the rest of the time interval. Thus, the parameter $\tau_c$ sets the velocity correlation time.
The velocity field is chosen to be a Couette
flow, \textit{i.e.} a two-dimensional linear shear flow, with a direction that is randomly rotated by an angle $\theta_n$
at the beginning of each time interval $\mathscr{I}_n$ \citep{y99,y09}.
For $t\in\mathscr{I}_n$ the velocity gradient takes the form
\begin{equation}\label{eq:kapparenew}
\bm\kappa=\sigma\begin{pmatrix}
-\sin\theta_n\cos\theta_n & \cos^2\theta_n
\\
-\sin^2\theta_n & \sin\theta_n\cos\theta_n
\end{pmatrix},
\end{equation}
where $\sigma$ is the magnitude of the shear and $\theta_n$ is distributed uniformly over $[0,2\pi]$.
The Lyapunov exponent as well as the generalized Lyapunov
exponents for this flow can be calculated exactly~\citep{y99,y09}:
\begin{equation}\label{eq:lambdarenew}
\lambda = \frac{1}{2\tau_c}\,\ln\left(1+\frac{\sigma^2\tau_c^2}{4}\right),
\qquad
\mathscr{L}(q)=\frac{1}{\tau_c}\ln\left[P_{q/2}\left(1+\frac{\sigma^2\tau_c^2}{2}\right)\right],
\end{equation}
where $P_{q/2}$ is the Legendre function of order $q/2$.

The solution of \eqref{eq:alpha} for the renewing Couette flow is presented in figure~\ref{fig:couette}(\textit{a}) for several values of $\sigma\tau_c$ (this non-dimensional group is the only free parameter that remains after substituting \eqref{eq:lambdarenew} in \eqref{eq:alpha}). As $\tau_c$ is decreased towards zero the results are seen to approach the prediction for a delta-correlated flow~\eqref{eq:alpha-kraichnan-d}, shown by the dashed (black) line. We also expect the results for all cases of $\sigma\tau_c$ to be well approximated by~\eqref{eq:alpha-kraichnan-d} in the vicinity of the coil--stretch transition, $\Wi = \Wicr = 1/2$. This is indeed the case. In addition, for this renewing flow, ~\eqref{eq:alpha-kraichnan-d} is seen to provide an excellent approximation for all $\Wi$ greater than $\Wicr$. In general, one may expect the deviation of $\alpha$ from the prediction of \eqref{eq:alpha-kraichnan-d} to be higher for small $\Wi$ than for large $\Wi$. This is because $\alpha$ is negative for large $\Wi$ and the form of $\Lscr(q)$ for negative arguments is strongly constrained by its convexity and the general properties $\Lscr(0)=\Lscr(-d)=0$ and $\Lscr'(0)=\lambda$.

In figure~\ref{fig:couette}(\textit{b}) the prediction of  \eqref{eq:alpha} is compared with the results of Brownian dynamics (BD) simulations of the dumbbell model (\S~\ref{sec:dumbbell}), with $\bm\kappa$ given by \eqref{eq:kapparenew}, for the case of $\tau_c = 0.1$ and $\sigma=10$. The decorrelated-flow approximation is also shown for comparison (black dashed line).
To obtain $\alpha$ from the simulations, we fit the stationary p.d.f of the extension
$p(R)$ with a power law over a range $\Req\ll R\ll \Rmax$. Figure~\ref{fig:couette}(\textit{b}) shows excellent agreement between the
large deviations theory and the simulations of the dumbbell model. 

It is interesting to note that \eqref{eq:alpha} may also be regarded as a tool for measuring the generalized Lyapunov exponents of
a turbulent flow from the statistics of polymer extensions. Since $\alpha$ is a monotonic function of $\Wi$, 
one could invert the $\alpha$ vs $\Wi$ relation obtained from simulations and express $\Wi$ in terms of $\alpha$ on the left-hand-side of \eqref{eq:alpha}, so as to obtain an explicit formula for $\mathscr{L}(\alpha)$. However, even with this strategy, measuring the generalized
Lyapunov exponents for large negative and positive orders will remain challenging. On the one hand, because $\alpha\geq -d$ this approach cannot yield the generalized Lyapunov exponents of order
less than $-d$. On the other hand, it is computationally intensive to construct the tail of the p.d.f. of $R$ for small $\Wi$, which corresponds to large positive values of $\alpha$ and hence to generalized Lyapunov exponents of large positive order.

\subsection{Stretching in turbulence and the roles of mild and extreme strain rates}
\label{sec:Gaussian}

\begin{figure}
\centering
\begin{overpic}[width=.9\textwidth]{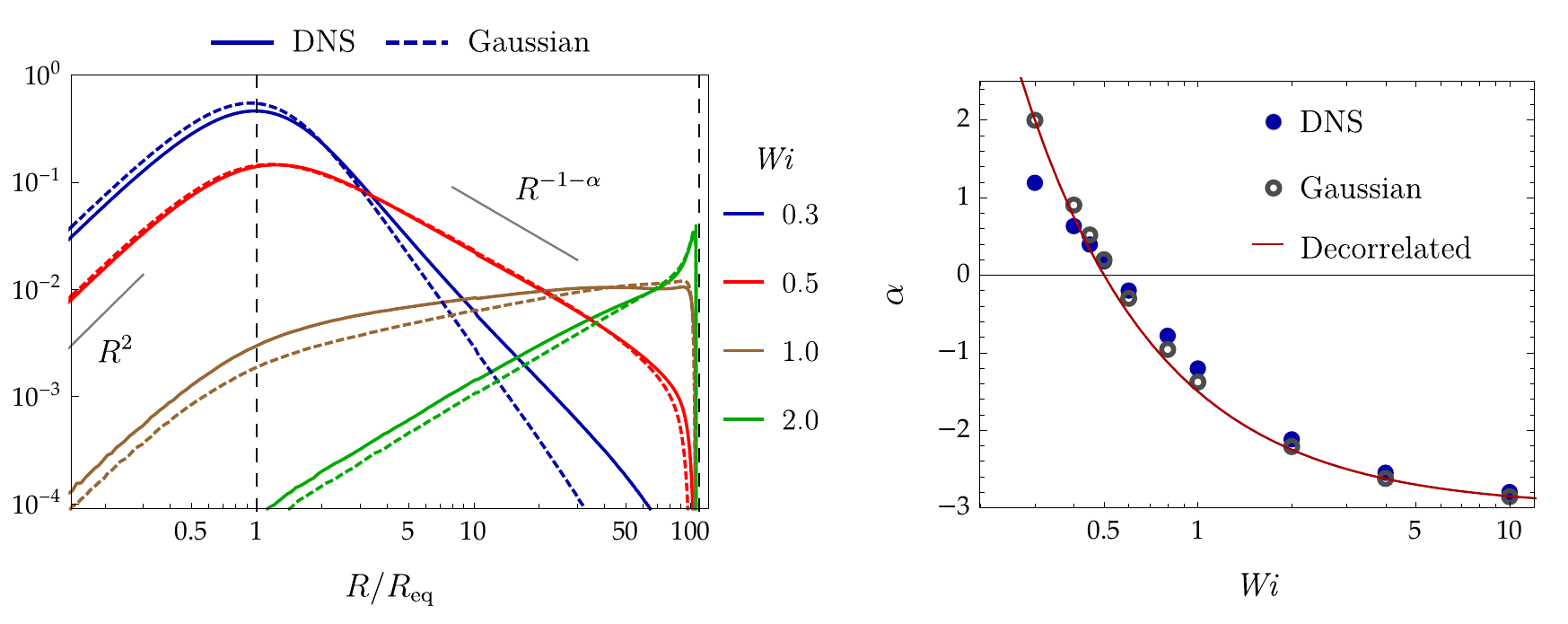}
\put (-3,35) {(\textit{a})}
\put (56,35) {(\textit{b})}
\end{overpic}
\caption{Stationary probability distributions functions of the end-to-end extension $R$ of polymers in turbulent and random flows. (\textit{a}) Comparison of the stationary p.d.f. of $R$, for different values of $\Wi$, in a DNS of turbulent flow and a synthetic Gaussian flow, constructed with the same Lyapunov exponent $\lambda$ as the DNS, as well as the same Lagrangian correlation times for vorticity, $\tau_\Omega$, and strain rate, $\tau_S$. (\textit{b}) The power-law exponent of the tail of the p.d.f. of $R$ (panel \textit{a}) presented as a function of $\Wi$. The exponents from DNS are compared with that from the Gaussian flow, as well as with the prediction for a three-dimensional time-decorrelated random flow in \eqref{eq:alpha-kraichnan-d}.}
\label{fig:powF}
\end{figure}

Let us now examine the p.d.f. of the extension for FENE dumbbells in homogeneous isotropic turbulence (\S~\ref{sec:DNStraj}). These results are presented in figure~\ref{fig:powF}(\textit{a}) (solid lines). A power-law range is apparent for $R$ between $\Req$ and $\Rmax$ (vertical dashed lines) and the corresponding exponents, $-1-\alpha$, are seen to increase past $-1$ as $\Wi$ increases beyond $\Wi_\mathrm{cr}$. (The $R^2$ behaviour for $R<\Req$ is a consequence of thermal fluctuations.) This panel also shows results for a Gaussian random flow (dotted lines), constructed so as to match the turbulent flow in terms of its integral correlation times of vorticity and strain as well as its Lyapunov exponent $\lambda$ (see~\S~\ref{sec:Gaussian-grad}). Of course, the higher-order generalized Lyapunov exponents of these two flows will not be the same \citep*[see][]{bmv14}, and this is reflected in the differing values of $\alpha$ shown in figure~\ref{fig:powF}(\textit{b}) (markers). The thin solid line in this panel corresponds to the result \eqref{eq:alpha-kraichnan-d} for a three-dimensional ($d = 3$) time-decorrelated random flow. Though small, the differences between these results show a systematic dependence on $\Wi$ which warrants further scrutiny.


 Consider first the effect of the non-Gaussian statistics of turbulence. Figure~\ref{fig:powF}(\textit{b}) shows that low-$\Wi$ stiff polymers stretch more in a turbulent flow than in a Gaussian flow: the values of $\alpha$ are less negative in the turbulent flow (compare the filled and open markers) which implies a greater power-law exponent for $p(R)$. Therefore, encountering a higher frequency of extreme-valued velocity gradients aids in stretching stiff polymers. 
 Surprisingly, this is no longer true when $\Wi$ is increased beyond $\Wicr$. Rather, these moderately high-$\Wi$ polymers which are relatively easy to stretch are seen to be more extended in a Gaussian flow than in a turbulent flow. On further increasing $\Wi$, all three flows in figure~\ref{fig:powF}(\textit{b}) eventually exhibit nearly identical values of $\alpha$; this is to be expected since $\alpha$ must attain the limiting value of $-3$ regardless of flow statistics (see \S~\ref{sec:renewing}).

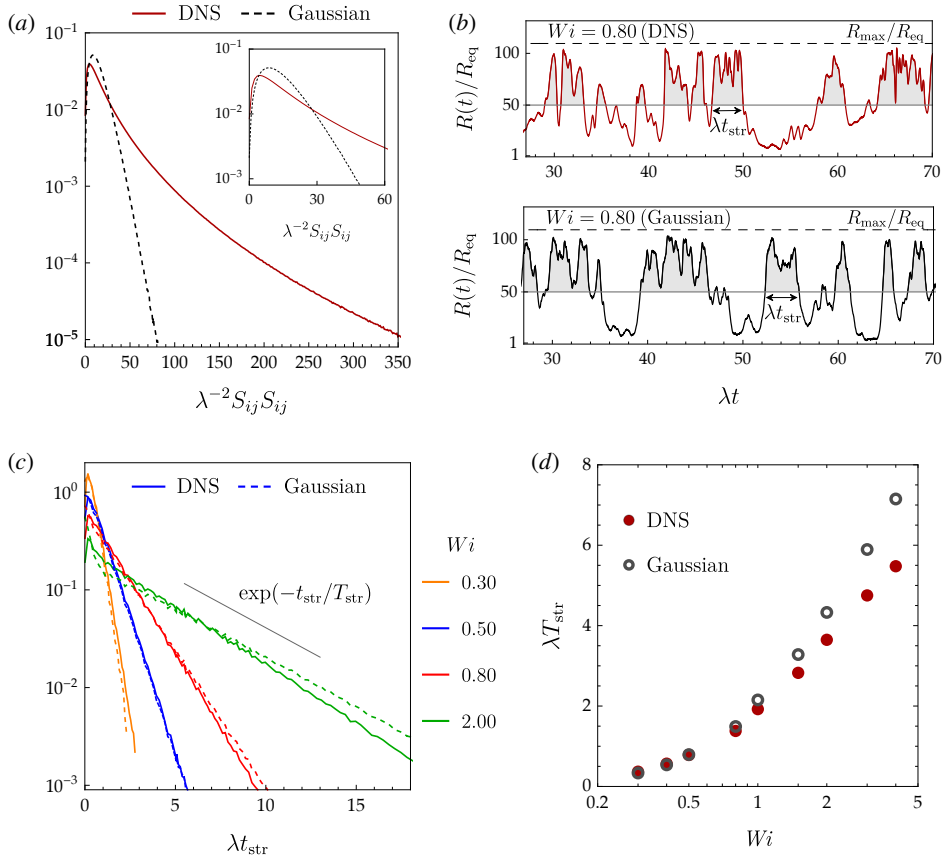
\begin{figure}
\centering
\begin{overpic}[width=.9\textwidth]{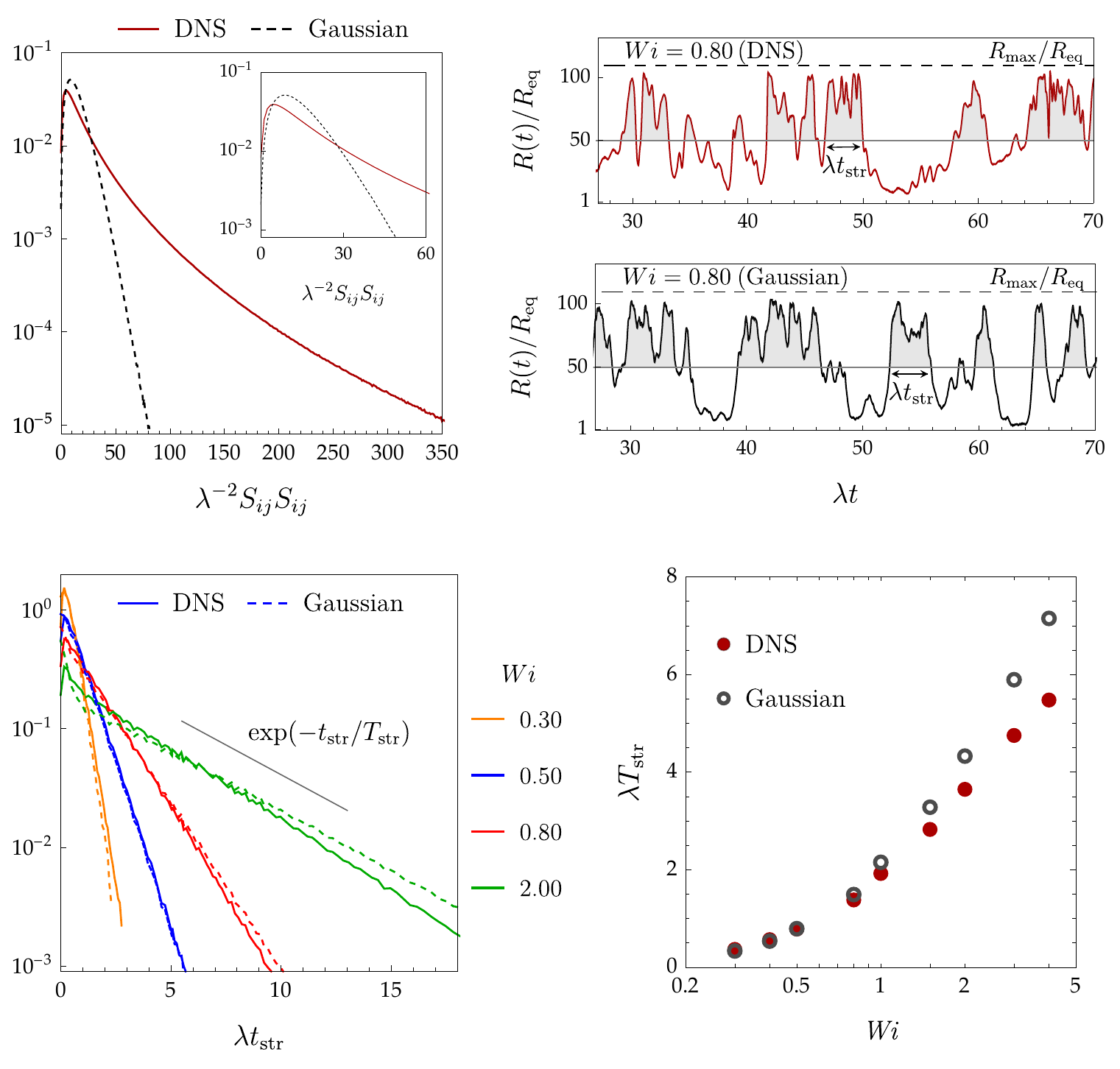}
\put (-3,91) {(\textit{a})}
\put (45,91) {(\textit{b})}
\put (-3,43.5) {(\textit{c})}
\put (54,43.5) {(\textit{d})}
\end{overpic}
\caption{Effect of non-Gaussian velocity-gradient fluctuations on polymer stretching in turbulence. (\textit{a}) Comparison of the p.d.f. of the strain rate sampled by polymers in a DNS of turbulent flow with that in a synthetic Gaussian flow with same $\lambda$, $\tau_\Omega$, $\tau_S$ as the DNS. The inset is a zoom which shows the near-peak behaviour of the distributions. (\textit{b})~Illustration of the typical stretching dynamics of a $\Wi = 0.8$ polymer in the DNS (top panel) and Gaussian flow (bottom panel). The grey shading shows when the polymer is stretched beyond a threshold of $\ell = \Rmax/2=50$; each such time interval yields a value of $t_{\mathrm str}$. (\textit{c}) Distributions of $t_{\mathrm str}$ for various values of $\Wi$ in both DNS and Gaussian flows. The exponential tails of these p.d.f.s are associated with the time-scale $T_{\mathrm str}$. (\textit{d}) Variation of $T_{\mathrm str}$, the typical time spent by polymers in a stretched state, as a function of $\Wi$, for both DNS and Gaussian flows. High-$\Wi$ polymers in the Gaussian flow are seen to persist in  a stretched state for significantly longer than they do in the DNS.}
\label{fig:persist}
\end{figure}

Why are moderately high-$\Wi$ polymers stretched more in a Gaussian flow? To answer this, it is helpful to examine the p.d.f. of the rate of strain $s = \sqrt{S_{ij} S_{ij}}$ sampled by polymers. Figure~\ref{fig:persist}(\textit{a}) compares the distributions of $s$ for the turbulent and Gaussian flows. We see that the Gaussian flow has a comparative abundance of mild strain-rate events to compensate for its lack of extreme-valued events. This suggests that high-$\Wi$ polymers that have long relaxation times are stretched more effectively by mild persistent straining rather than by strong but short-lived straining. In contrast, stretching small-$\Wi$ polymers which have short relaxation times requires strong strain-rate events. 
These observations are 
consistent with a previous result by \citet{tdmsl04} for a turbulent channel flow. It was shown that stretching events at low \textit{Wi} are typically preceded by a burst of the strain rate; such bursts were not seen at high \textit{Wi}.

If it is true that high-$\Wi$ polymers are stretched primarily by mild persistent straining, then they should not only stretch more in a Gaussian flow but also remain in an extended configuration for a longer duration of time. To detect this behaviour, we carry out a persistence time analysis and quantitatively compare how long polymers stay stretched in the turbulent and Gaussian random flows. Interestingly, the concept of persistence, which arose out of problems in non-equilibrium statistical physics~\citep{Satya-Review,Bray}, has recently been used to study the turbulent transport of particles and filaments (\citealt{Prasad,Bos,Akshay}; \citealt*{settling}).

We begin by defining a polymer to be in a `stretched' state if $R>\ell$, where the threshold $\ell = \Rmax/2$ is set well-within the power-law range. The non-stretched state is then defined as $  R \le \ell$. We have verified that varying the value of $\ell$ within the power-law range does not change our conclusions. With these states defined, we examine the Lagrangian history of each polymer and detect the time intervals $t_\mathrm{str}$ over which the polymer remains in a stretched state (see figure~\ref{fig:persist}\textit{b}). The distribution of this persistence time, $p(t_\mathrm{str})$, is presented in figure~\ref{fig:persist}(\textit{c}) for various $\Wi$ and for both the turbulent and Gaussian flows. At large $t_\mathrm{str}$ the distribution displays an exponential tail, $p(t_\mathrm{str}) \sim \mathrm{exp}(-t_\mathrm{str}/T_\mathrm{str})$, from which we extract the persistence time scale $T_\mathrm{str}$. 

Figure~\ref{fig:persist}(\textit{d}) presents $T_\mathrm{str}$ for both flows and for various values of $\Wi$. Clearly, polymers with $\Wi > \Wicr$ typically remain stretched for a significantly longer time in the Gaussian flow as compared to the turbulent flow. This is true even for very large $Wi$ for which the exponent of the power-law of $p(R)$ is nearly the same in both flows (see the results for $\Wi \ge 2$ in figures~\ref{fig:powF}\textit{b}  and~\ref{fig:persist}\textit{d}). So though the probability of large extensions is the same, the nature of stretching is different: Polymers experience many short-lived episodes of large extension in the turbulent flow, whereas such episodes are fewer but last longer in the Gaussian flow.

\section{Temporal evolution of the distribution of polymer extensions}
\label{sec:equilibration}

\subsection{Two regimes of evolution}

Thus far we have been studying the stationary p.d.f. of $R$, attained after the polymers have spent enough time in the flow for their statistics to equilibrate. We now examine how the p.d.f. of $R$ evolves with time. Past work has shown that the p.d.f. relaxes exponentially to its stationary form, with a $\Wi$-dependent time scale that exhibits a pronounced maximum at the coil--stretch transition~\citep{cpv06,wg10}. But what is the shape of the p.d.f. as it evolves? And how, if at all, does this evolution depend on $\Wi$? 

To answer these questions, we use our Lagrangian simulations to construct the p.d.f. of $R$ as a function of time, $p(R,t)$. We consider an initial state in which the polymers are in equilibrium with a static fluid. And so we first evolve the polymers with $\bm\kappa = 0$ (in~\eqref{eq:dumbbell}) until the p.d.f. of $R$ attains the equilibrium distribution \citep{bird}. We then `turn on' the turbulent flow.

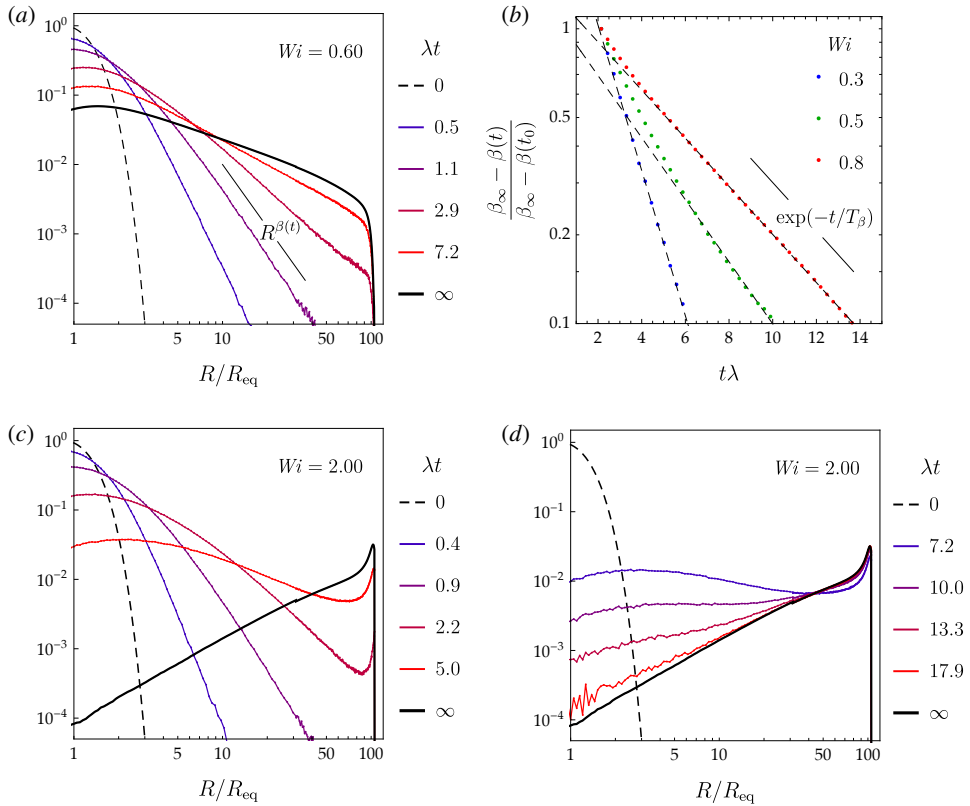
\begin{figure}
\centering
\begin{overpic}[width=.93\textwidth]{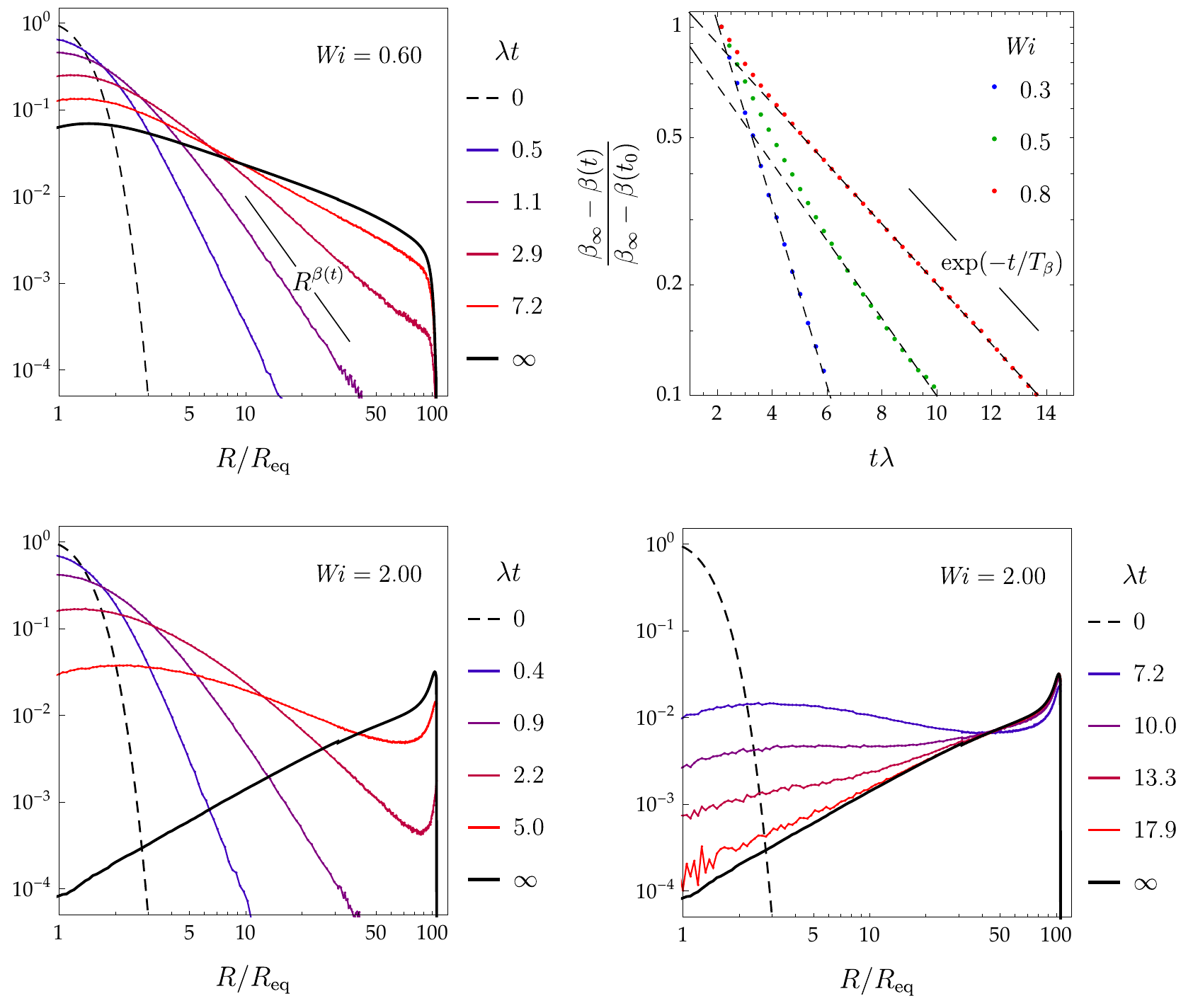}
\put (-2,83) {(\textit{a})}
\put (50,83) {(\textit{b})}
\put (-2,39) {(\textit{c})}
\put (50,39) {(\textit{d})}
\end{overpic}
\caption{Two regimes of evolution of the p.d.f. of the polymer extension. (\textit{a}) Depiction of the\textit{ evolving power-law} regime, seen for small to moderate $\Wi$, in which the tail of $p(R,t)$ evolves as $R^{\beta(t)}$. The thick black line with the time stamp $\lambda t = \infty$ represents the stationary p.d.f. of $R$. (\textit{b}) The power-law exponent $\beta(t)$ is seen to approach its steady state value $\beta_\infty = -1-\alpha$ exponentially. Here, $t_0$ is a reference time. (\textit{c})-(\textit{d}) Depiction of the high-$Wi$ \textit{rapid stretching} regime, in which $p(R,t)$ does not evolve as a power law; rather the p.d.f. quickly forms a local maximum near to $\Rmax$ (panel \textit{c}) and then adjusts its shape directly to that of the stationary power law (panel \textit{d}).}
\label{fig:time_DNS}
\end{figure}

The evolution of $p(R,t)$ is illustrated in figure~\ref{fig:time_DNS}. Interestingly, we find two qualitatively distinct regimes. At small or moderate $\Wi$ (figure~\ref{fig:time_DNS}\textit{a}), the p.d.f. is seen to quickly attain a power-law form with an exponent $\beta(t)$ that increases in time until it reaches its stationary value $\beta_\infty = -1-\alpha$. By fitting the distributions with a power law in the range $R_\mathrm{eq} \ll R \ll \Rmax$, we find that $\beta(t)$ relaxes exponentially, \textit{i.e.} $\beta_\infty-\beta \sim \mathrm{exp}(-t/T_\beta)$, as demonstrated in figure~\ref{fig:time_DNS}(\textit{b}). The time-scale $T_\beta$ is analyzed later in \S~\ref{sec:Tbeta}.

The evolution at high $\Wi$ is quite different and is shown in figure~\ref{fig:time_DNS}(\textit{c})-(\textit{d}). Here, the polymers stretch rapidly and quickly produce a local peak close to the maximum extension $\Rmax$. Thus, although a transient power law appears at very early times, it is quickly lost as the local peak at $\Rmax$ begins to dominate the distribution (figure~\ref{fig:time_DNS}\textit{c}). The long-time equilibration of the p.d.f. occurs by the peak near $\Rmax$ first approaching its stationary value; then the stationary power law $R^{\beta_\infty}$ gradually emerges, starting near $\Rmax$ and then extending its range down-scale towards $\Req$ (figure~\ref{fig:time_DNS}\textit{d}).

These two regimes of equilibration are termed the \textit{evolving power-law} regime and the \textit{rapid-stretching} regime. The former occurs for $\Wi \lesssim 3/4$, while the latter occurs for larger $\Wi$. Note that the cross-over point, $\Wi = 3/4$, is marked by a stationary p.d.f. with $\beta_\infty = 0$ that has neither a decaying tail at large $R$ nor a local peak near $\Rmax$; the evolution near $\Wi = 3/4$ is a blend of the two regimes.

\subsection{The time-dependent power-law: insights from a stochastic model} \label{sec:FP}

We now lend credence to the above characterization of the evolving power-law regime by using a stochastic model to show that, for $0\leq\Wi\lesssim 3/4$, an evolving power-law is a natural consequence of scale separation between $\Req$ and $\Rmax$. The associated analysis also reveals the $\Wi$ dependence of the relaxation time-scale $T_\beta$.

We consider the Batchelor--Kraichnan flow wherein the velocity gradient $\bm\kappa(t)$ is a statistically isotropic, time-decorrelated, $3\times 3$ Gaussian tensor \citep{fgv01}. In terms of the scaled variables $r = R/\Req$ and $\tilde{t}=t/2\tau$, the p.d.f. of the extension $p(r,\tilde{t})$ is governed by a Fokker--Planck equation \citep{mav05,pav16}:
\begin{equation}
\label{eq:FPE}
\partial_{\tilde{t}} p=-\partial_r[D_1(r)p]+\partial_r^2[D_2(r)p],
\end{equation}
where $f(r)=1/(1-r^2/r_m^2)$, $\rmax = \Rmax/\Req$, and
\begin{equation}
D_1(r)=[8\Wi/3-f(r)]r+2/(3r), \quad D_2(r)=2\Wi\, r^2/3+1/3.
\end{equation}
Note that while \eqref{eq:FPE} is exact for the Batchelor--Kraichnan flow, an equation of the same form
may be obtained for general turbulent flows by considering the Fokker--Planck equation associated with \eqref{eq:dumbbell},
assuming statistical isotropy, and modelling the stretching term \textit{\`a la} Richardson, \textit{i.e.} as a diffusive process. The associated {extension-dependent} eddy diffusivity should scale as $R^2$ since the extension remains below the viscous scale.

We assume a wide separation between the scales of the equilibrium and maximum extensions, i.e. $\Rmax \gg \Req$, or $\rmax \gg 1$. Further, we focus on the long-time evolution of $p(r,\tilde{t})$ which is characterized by distinct behaviours for small, intermediate, and very large extensions:
(i) In the range of extensions where thermal fluctuations dominate ($r<1$), the p.d.f. of $r$ assumes a quadratic profile (figure \ref{fig:powF});
(ii) For intermediate extensions $1\ll r\ll r_m$, a power-law solution $r^{\beta(\tilde{t})}$ forms with an exponent $\beta(\tilde{t})$ that converges to its stationary value $\beta_\infty$;
(iii) For extreme extensions, $p(r,\tilde{t})$ decreases rapidly as $r$ approaches $r_m$.
We therefore introduce the ansatz:
\begin{subnumcases}{p(r,\tilde{t}) \sim}
c(\tilde{t})\, r^2 &  $(0\leq r\leq 1)$,
\\
c(\tilde{t})\, r^{\beta(\tilde{t})} &  $(1< r < \ratio r_m)$,
\label{eq:ansatz}
\\
g(r,\tilde{t}) & $(\ratio r_m \leq r \leq r_m)$.
\end{subnumcases}
Here, $\ratio$ is a threshold value of $r/\rmax$, less than unity, beyond which the nonlinear part of the FENE spring coefficient $f(r)$ becomes non-negligible and terminates the power-law behaviour.
The function $g(r,\tilde{t})$ is such that, for $r = \ratio \rmax$, we have $g(\ratio \rmax,\tilde{t})=c(\tilde{t}) (\ratio \rmax)^{\beta(\tilde{t})}$; while for $r > \ratio \rmax$, $g(r,\tilde{t})$ decreases faster than $r^{\beta(\tilde{t})}$ as $r$ approaches $\rmax$.
The latter assumption only applies if $\Wi < 3/4$, for which $\beta_\infty < 0$. For larger $\Wi$, our Lagrangian simulations show that the p.d.f. of $r$
develops a pronounced time-dependent peak between $\gamma r_m$ and $r_m$ (see figures \ref{fig:time_DNS}\textit{c} and \ref{fig:time_DNS}\textit{d}). Thus, the subsequent analysis applies only for $\Wi < 3/4$, which in fact corresponds to the regime in which the Lagrangian simulations exhibit an evolving power law.

The coefficient $c(\tilde{t})$ is determined in terms of the exponent $\beta(\tilde{t})$ through the normalization condition
\begin{equation}
\label{eq:normalization}
1=
\underbracket{\int_{0}^1 c(\tilde{t})r^2  \mathrm{d}r}_{I_1} + 
\underbracket{\int_{1}^{\ratio\rmax} c(\tilde{t}) r^{\beta(\tilde{t})} \mathrm{d}r}_{I_2} + 
\underbracket{\int_{\ratio\rmax}^{\rmax} g(r,\tilde{t}) \mathrm{d}r}_{I_3}.
\end{equation}
Clearly, $I_1 = c/3$, while $I_2 = c[(\gamma r_m)^{1+\beta}-1]/[1+\beta]$ for $\beta \neq 1$ and $I_2 = c \ln(\ratio\rmax)$ for $\beta=-1$. For $r_m \gg 1$, $I_3$ is subdominant with respect to $I_1$ and $I_2$ and
will be ignored. 
Thus, solving \eqref{eq:normalization} for $c$ yields
\begin{subequations}
\begin{alignat}{2}
c &=  \frac{1+\beta}{(\ratio\rmax)^{1+\beta} + (\beta-2)/3}, &\quad& \text{for  $\beta \neq -1$},
\\ 
c &=  \frac{1}{1/3 +\ln(\ratio\rmax)}, && \text{for  $\beta = -1$}.
\end{alignat}
\label{eq:cfull}
\end{subequations}
To calculate $\beta(\tilde{t})$, we observe that, for $1 \ll r \ll \rmax$, \eqref{eq:FPE} can be simplified as follows: 
\begin{equation}
\label{eq:FPEapprox}
\partial_t p=-\partial_r\left[\left(\frac{8}{3}Wi-1 \right) r p\right]+\partial_r^2\left(\frac{2}{3}Wi \,r^2 p\right).
\end{equation}
At steady-state we know that $p(r) \propto r^{\beta_\infty}$, which when substituted into \eqref{eq:FPEapprox} yields
$\beta_\infty=2 -3/2\Wi$, in accordance with the large deviations theory for a time-decorrelated flow (see \eqref{eq:alpha-kraichnan-d} and recall that $\beta_\infty = -1-\alpha$).

Now, substituting $p$ from \eqref{eq:ansatz} into \eqref{eq:FPEapprox} results in
\begin{equation}
\label{eq:cbeta}
\left(\frac{\mathrm{d}c}{\mathrm{d} \beta}+c\ln r\right) \frac{\mathrm{d}\beta}{\mathrm{d}\tilde{t}}=c F(\beta,\Wi)
\end{equation}
with
\begin{equation}
F(\beta,\Wi)= -\left(\frac{8}{3}Wi-1 \right)(1+\beta)+\frac{2}{3}Wi(1+\beta)(2+\beta).
\end{equation}
The left hand side of \eqref{eq:cbeta} contains $\mathrm{d}c/{\mathrm{d} \beta}$ which is evaluated using \eqref{eq:cfull}:
\begin{subequations}
\begin{alignat}{2}
    \frac{\mathrm{d}c}{\mathrm{d} \beta} &= c\left[\frac{1}{1+\beta}-\frac{(\ratio\rmax)^{1+\beta} \ln{(\ratio\rmax)}+1/3}{(\ratio\rmax)^{1+\beta} +(\beta-2)/3}\right], &\quad& \text{for $\beta \neq -1$}, \label{eq:dcdba}\\
    \frac{\mathrm{d}c}{\mathrm{d} \beta} &=0, && \text{for $\beta = -1$}.
\end{alignat}
\end{subequations}
On taking the limit $\rmax \gg 1$, the right-hand-side of \eqref{eq:dcdba} simplifies to $-3c/(\beta^2-\beta-2)$ for $\beta < -1$ and to $-c\ln{(\ratio\rmax)}$ for $\beta > -1$. By substituting these expressions for $\mathrm{d}c/{\mathrm{d} \beta}$ in \eqref{eq:cbeta} and considering that $\rmax \gg r\gg 1$, we obtain the following leading-order equation for $\beta$:
\begin{subequations}
\label{eq:betaeq}
\begin{alignat}{2}
\ln(r)\, \frac{\mathrm{d}\beta}{\mathrm{d}\tilde{t}}&= F(\beta,\Wi), &\quad& \text{for $\beta \leq -1$}, \label{eq:betaeqa}\\
\ln(\ratio r_m)\; \frac{\mathrm{d}\beta}{\mathrm{d}\tilde{t}}&= -F(\beta,\Wi), && \text{for $\beta > -1$}. \label{eq:betaeqb}
\end{alignat}
\end{subequations}
Because $\ln{(x)}$ is a weak function of $x$ for $x\gg 1$, we approximate $\ln r$ in \eqref{eq:betaeqa}, as well as $\ln(\ratio r_m)$ in \eqref{eq:betaeqb}, as $\ln r_m$ to obtain
 \begin{equation}\label{eq:approxpow}
 \frac{\mathrm{d}\beta}{\mathrm{d}\tilde{t}} \approx a\; \frac{F(\beta,\Wi)}{\ln r_m},
 \end{equation}
with $a = 1$ for $\beta \leq -1$ and $a=-1$ for $\beta > -1$. 

Being independent of $r$, \eqref{eq:approxpow} shows that $p(R,t)$ may be approximated (up to logarithmic corrections) by a time-dependent power-law in the range $\Req \ll R \ll \Rmax$. 

Next, to reveal the long-time relaxation of $\beta$ to $\beta_\infty$, we substitute $\beta = \beta_{\infty}+ \beta'$ in \eqref{eq:approxpow}
and linearize for small $\beta'$. Noting that $F(\beta_{\infty},Wi) = 0$ and that $\beta_{\infty} < -1$ for $\Wi < Wi_\mathrm{cr}$, we obtain
\begin{subequations}\label{eq:betap}
\begin{align}
\frac{\mathrm{d}\beta'}{\mathrm{d}\tilde{t}} &= -\frac{2 \left({\Wicr-\Wi}\right)}{\ln r_m}\beta', \quad {\rm for}\;\;\Wi < \Wicr,\\
\frac{\mathrm{d}\beta'}{\mathrm{d}\tilde{t}} &= -\frac{2 \left({\Wi-\Wicr}\right)}{\ln r_m}\beta', \quad {\rm for}\;\;\Wicr < \Wi < 3/4.
\end{align}
\end{subequations}
In terms of dimensional time $t$, this implies an exponential relaxation of the power-law exponent,
\begin{equation}\label{eq:exprelax}
\beta_{\infty} - \beta(t) \sim {\rm exp}(-t/T_\beta),
\end{equation}
with a time scale 
\begin{equation}\label{eq:Widiv}
T_\beta \sim \frac{\tau}{|\Wi-\Wicr|}
\end{equation}
that diverges as $\Wi$ approaches $\Wi_\mathrm{cr}$.

Note that \eqref{eq:approxpow} actually has two fixed points: $\beta_\infty$ and $-1$. The latter, though, can be shown to be dynamically unstable, leaving $\beta_\infty$ as the only stable fixed point.

\begin{figure}
\centering
\begin{overpic}[width=1\textwidth]{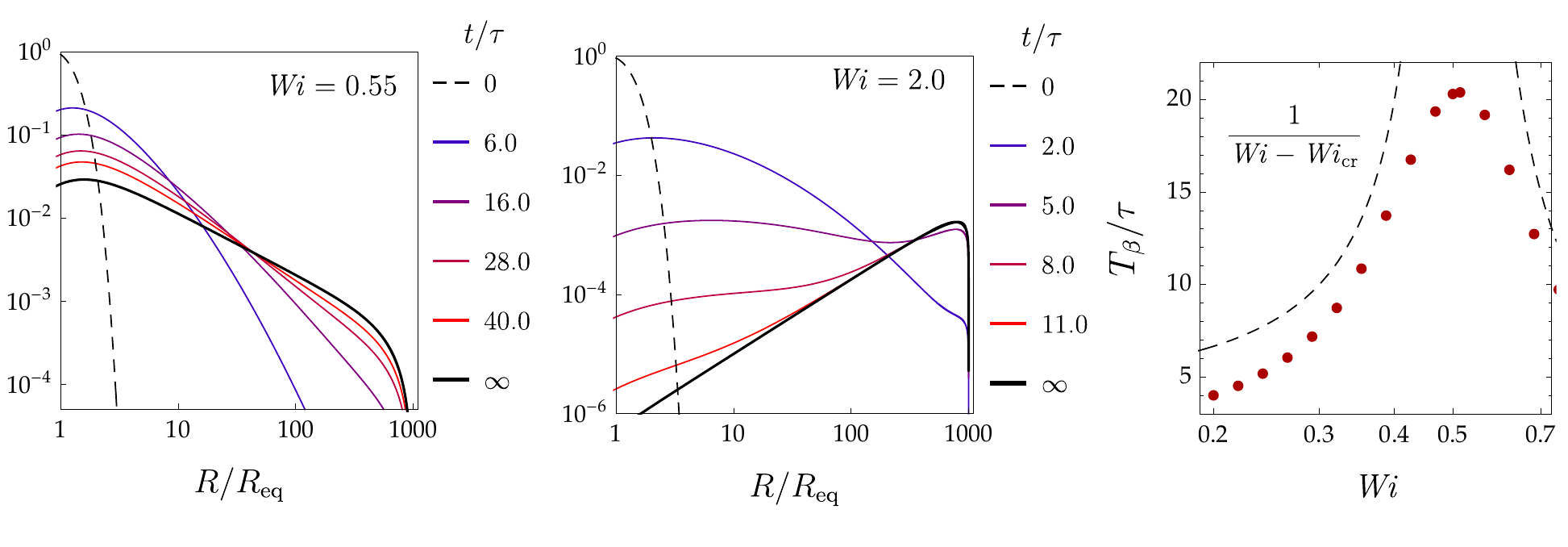}
\put (-2,31.5) {(\textit{a})}
\put (33.5,31.5) {(\textit{b})}
\put (71,31.5) {(\textit{c})}
\end{overpic}
\caption{Evolution of the p.d.f. of the extension as predicted by numerical simulations of the Fokker--Planck equation for the Batchelor--Kraichnan flow (with $\Rmax = 10^3 \Req$). The evolving power-law regime is shown in panel (\textit{a}), while the rapid-stretching regime is shown in panel (\textit{b}). The time-scale of relaxation $T_\beta$ of the evolving power-law exponent is shown in panel (\textit{c}) as a function of $\Wi$. The dashed line shows the variation predicted by the asymptotic analysis [cf. \eqref{eq:exprelax}].}
\label{fig:FP}
\end{figure}

The prediction \eqref{eq:exprelax} of an exponentially relaxing power-law exponent is certainly in agreement with the results of our Lagrangian simulations (see figure~\ref{fig:time_DNS}\textit{b}). The $\Wi$-dependence of the relaxation time-scale will be examined in light of \eqref{eq:Widiv} in the next section. But first, it is instructive to compare  \eqref{eq:exprelax}-\eqref{eq:Widiv} against
numerical simulations of the Fokker-Planck equation \eqref{eq:FPE}. Beginning from a distribution of coiled dumbbells, we solve \eqref{eq:FPE} using second-order central differences and the LSODA algorithm \citep{lsodaref} with adaptive time-stepping \citep{ndsolve}. We take $\Rmax = 10^3 \Req$, thereby realizing a much larger scale separation than is practical in our Lagrangian simulations. The results are presented in figure~\ref{fig:FP}. 
%

The stochastic model exhibits the same two regimes of evolution as the Lagrangian simulations: the evolving power-law regimes for $\Wi < 3/4$ (cf. figures \ref{fig:time_DNS}\textit{a} and \ref{fig:FP}\textit{a}) and the rapid-stretching regime for larger $\Wi$ (cf. figures \ref{fig:time_DNS}\textit{c}-\ref{fig:time_DNS}\textit{d} and \ref{fig:FP}\textit{b}). 
Within the evolving power-law regime, we extract the exponent $\beta(t)$ and find that it does in fact evolve exponentially, in accordance with \eqref{eq:exprelax}. The time-scale $T_\beta$, obtained from a least-squares fit, is presented as a function of $\Wi$ in figure~\ref{fig:FP}(\textit{c}). The predicted divergence of $T_\beta$ at $\Wicr$ (see \eqref{eq:Widiv}), manifests in the simulations (which have a finite scale separation) as a prominent maximum of $T_\beta$ at $\Wicr$. Such a slowing down of the stretching dynamics at the coil--stretch transition was demonstrated by \citet{mav05} and \citet{cpv06}, but in regard to the equilibration of the entire p.d.f. of $R$. The behaviour of $T_\beta$ vs $\Wi$ gives an alternative characterization of the same phenomenon.

\subsection{Temporal correlation and non-Gaussian statistics slow the equilibration of $p(R,t)$}\label{sec:Tbeta}

We now quantitatively compare the time scales of equilibration of $p(R,t)$ in the turbulent and Gaussian flows (Lagrangian simulations) as well as in the Batchelor--Kraichnan decorrelated flow (numerical solution of the stochastic model).

One time scale of interest is that associated with the evolving power-law tail, $T_\beta$. However, this time scale is only relevant for $\Wi < 3/4$. So, to characterize the equilibration time for all $\Wi$, we use the time scale associated with the entire p.d.f. of $R$. \citet{mav05} and \citet{cpv06} showed, for random flows, that $p(R,t)$ approaches its asymptotic stationary form exponentially, 
with a time-scale $T_P$ that displays a maximum near $\Wicr$. \citet{wg10} confirmed this prediction in a DNS of isotropic turbulence. We obtain $T_P$ from our simulations by fitting the evolution of the first moment of $p(R,t)$ (approximately the same value is obtained by fitting the higher moments). 

\begin{figure}
\centering
\begin{overpic}[width=.85\textwidth]{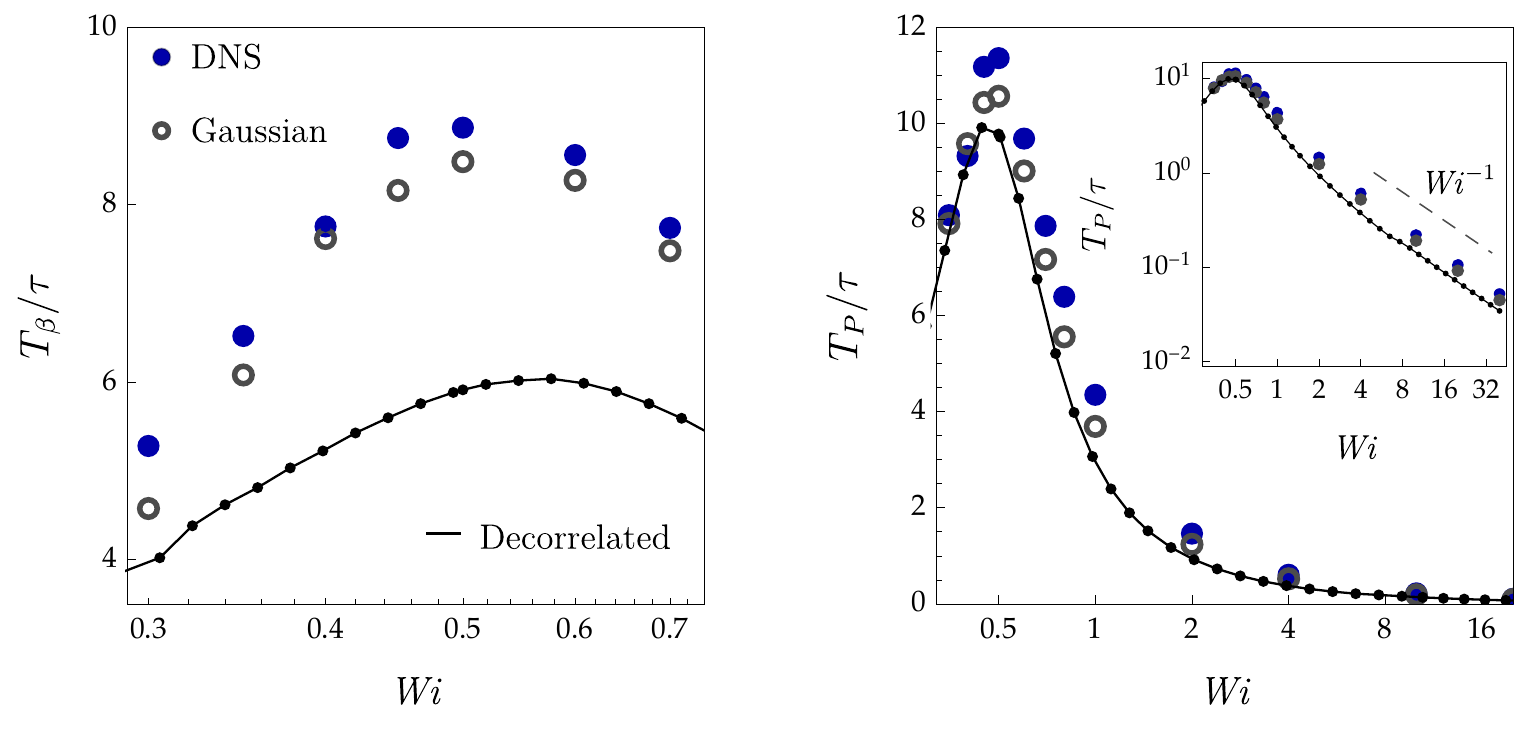}
\put (1,45.) {(\textit{a})}
\put (54,45) {(\textit{b})}
\end{overpic}
\caption{Effect of non-Gaussian velocity-gradient statistics on the time scale of evolution of the p.d.f. of the extension to its stationary form. (\textit{a}) Comparison of the relaxation time scale $T_\beta$ of the power-law exponent of the tail of the p.d.f. of $R$ obtained from DNS (cf. figure~\ref{fig:time_DNS}\textit{b}) with that from the synthetic Gaussian flow as well as the Batchelor--Kraichnan flow (simulations of \eqref{eq:FPE}). (\textit{b}) Comparison of the exponential relaxation time scale $T_P$ of the entire p.d.f. of $R$, for the three different cases: DNS, Gaussian flow, and Batchelor--Kraichnan flow. The inset is a log-log plot of the same data which shows the $Wi^{-1}$ asymptotic behaviour. }
\label{fig:timecomp}
\end{figure}

Figure~\ref{fig:timecomp} compares the equilibration time scales $T_\beta$ and $T_P$ in panels (\textit{a}) and (\textit{b}), respectively, for the turbulent flow (filled markers) and the Gaussian flow (open markers). The results for the time-decorrelated Batchelor--Kraichnan flow are also presented (line). We see that the qualitative variation of these time scales with $\Wi$ is the same in all flows and exhibits a maximum near $\Wicr$. In the case of $T_P$, a $\Wi^{-1}$ asymptotic behaviour is visible (see the inset of figure~\ref{fig:timecomp}\textit{b}), in agreement with \citet{mav05} and \citet{cpv06}.

On comparing the results for the Gaussian and decorrelated flows, we see that the time correlation of the flow slows down the equilibration of the p.d.f. of $R$: both $T_\beta$ and $T_P$ are higher for the Gaussian flow (figures~\ref{fig:timecomp}\textit{a} and ~\ref{fig:timecomp}\textit{b}). The equilibration is even slower for the turbulent flow (DNS) with its non-Gaussian velocity gradient statistics. 

Qualitatively, chaotic random flows are seen to provide a good approximation for studying polymer stretching in homogeneous isotropic turbulence. Our results also show that the accuracy of the predictions, especially for finite-time statistics, are improved by incorporating temporal correlation in the random flow. 

\section{Chains stretch like equivalent dumbbells}
\label{sec:chain}

Thus far we have focused on the simplest model for a polymer of finite extension, the FENE dumbbell (\S~\ref{sec:dumbbell}). We now show that the key results presented above also hold for a polymer chain, which incorporates higher-order deformation modes.

The three-dimensional motion of a freely-jointed (Rouse) chain, composed of $\mathscr{N}$ beads, is described
in terms of the position of its center of mass, $\bm X_c$, and the separation
vectors between the beads, $\bm Q_i$ ($i=1,\dots,\mathscr{N}-1$) \citep{bird,o96}:%
\begin{subequations}%
\begin{eqnarray}%
  \label{eq:cm}
  \dot{\bm X}_c &=& \bm u(\bm X_c(t),t)+\dfrac{1}{\mathscr{N}}\sqrt{\frac{\Qeq^2}{6\tau_*}}
  \sum_{i=1}^{\mathscr{N}}\bm\xi_i(t),
  \\[2mm]
  \label{eq:q}
  \dot{\bm Q_i} &=& 
  \bm\kappa(t)\bcdot\bm Q_i(t)-\dfrac{1}{4\tau_*}
           [2f_i\bm Q_i(t)-f_{i+1}\bm Q_{i+1}(t)-f_{i-1}\bm Q_{i-1}(t)]
           \\
           \nonumber
           &&+\sqrt{\dfrac{\Qeq^2}{6\tau_*}}
             [\bm\xi_{i+1}(t)-\bm\xi_i(t)],\qquad i=1,\dots,\mathscr{N}-1,%
\end{eqnarray}%
\label{eq:chain}%
\end{subequations}%
where each link is associated with an elastic time scale
$\tau_*$ and has an equilibrium r.m.s.
extension $\Qeq=\sqrt{3k_BT/H}$.
The FENE interactions between neighbouring beads is 
characterized by the coefficients
\begin{equation}
f_i=\dfrac{1}{1-\vert\bm Q_i\vert^2/Q_m^2},
\label{eqn:fi}
\end{equation}
which
ensure that the extension of each spring 
does not exceed its maximum length $Q_m$. The Brownian forces that act on the beads are
represented by independent, vectorial, white noises $\bm\xi_i(t)$. 
Note that one must set $\bm Q_0=\bm Q_{\mathscr{N}}=0$
in the equations for $\bm Q_1$ and $\bm Q_{\mathscr{N}-1}$.

The end-to-end separation or extension vector of the polymer chain is
defined as $\bm R=\sum_{i=1}^{\mathscr{N}-1}\bm Q_i$. In a still fluid, the equilibrium r.m.s. value of $\vert\bm R\vert$ is
$\Qeq\sqrt{\mathscr{N}-1}$ \citep{bird}. 

\begin{figure}
\centering
\begin{overpic}[width=.93\textwidth]{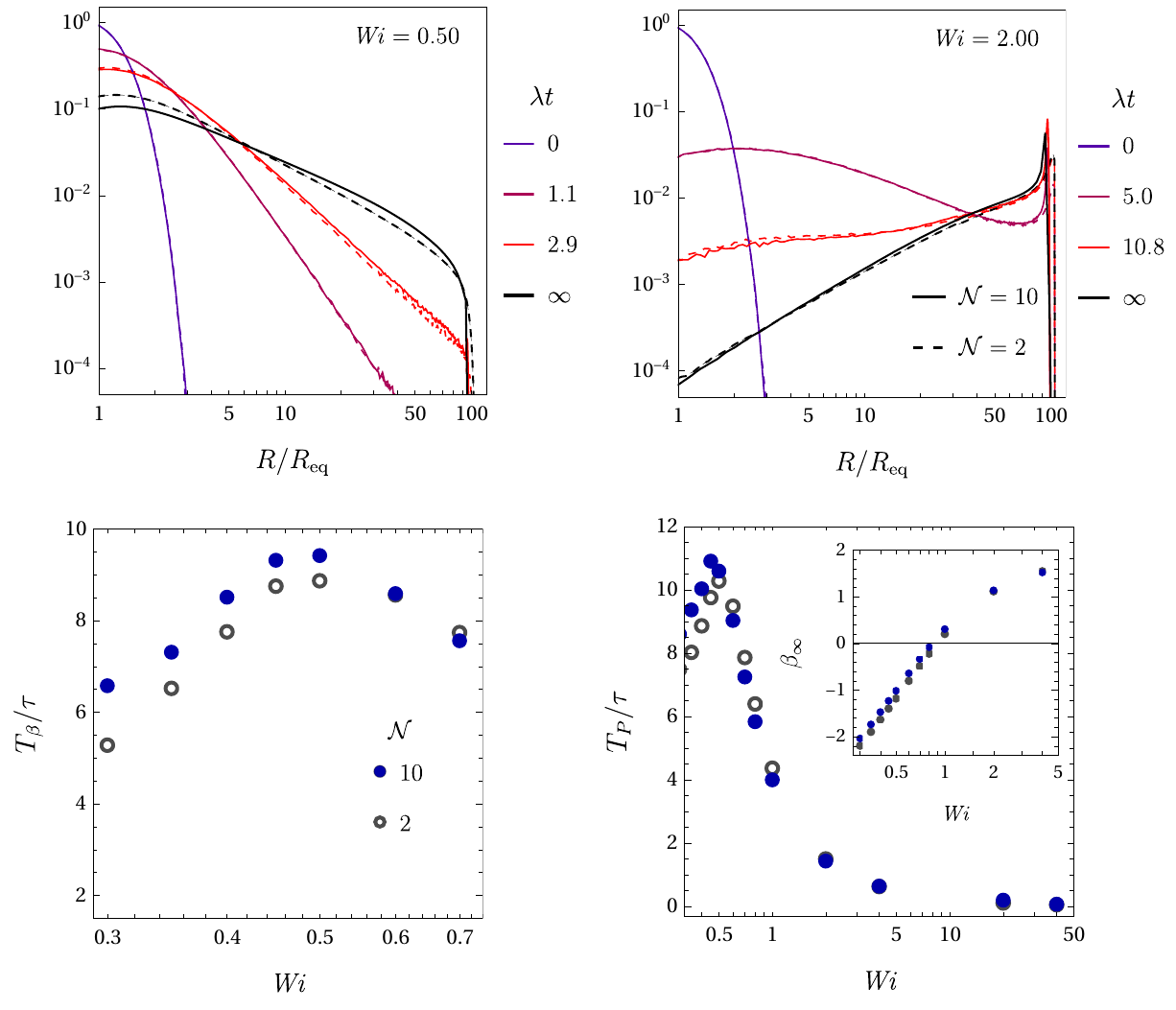}
\put (1,84) {(\textit{a})}
\put (50,84) {(\textit{b})}
\put (1,40) {(\textit{c})}
\put (50,40) {(\textit{d})}
\end{overpic}
\caption{Effect of the polymer model---bead-spring dumbbell or chain---on the evolution of the p.d.f. of the extension. (\textit{a}) Comparison of the evolving p.d.f. of $R$, in the evolving power-law regime, of a $\mathscr{N}=10$ chain (solid line) with that of an equivalent dumbbell (dashed line). The parameters of the dumbbell are given by \eqref{eq:jcmap}. (\textit{b}) Comparison of the evolution of the p.d.f. of $R$ in the rapid stretching regime. (\textit{c}) Comparison of the relaxation time $T_\beta$ of the power-law exponent of the evolving p.d.f. for the chain and dumbbell and for various $\Wi$ in the evolving-power-law regime. (\textit{d}) Comparison of the exponential relaxation time-scale $T_P$ of the entire p.d.f. of $R$ for the chain and the dumbbell. The inset compares the power-law exponents of the stationary p.d.f.s.}
\label{fig:chain}
\end{figure}
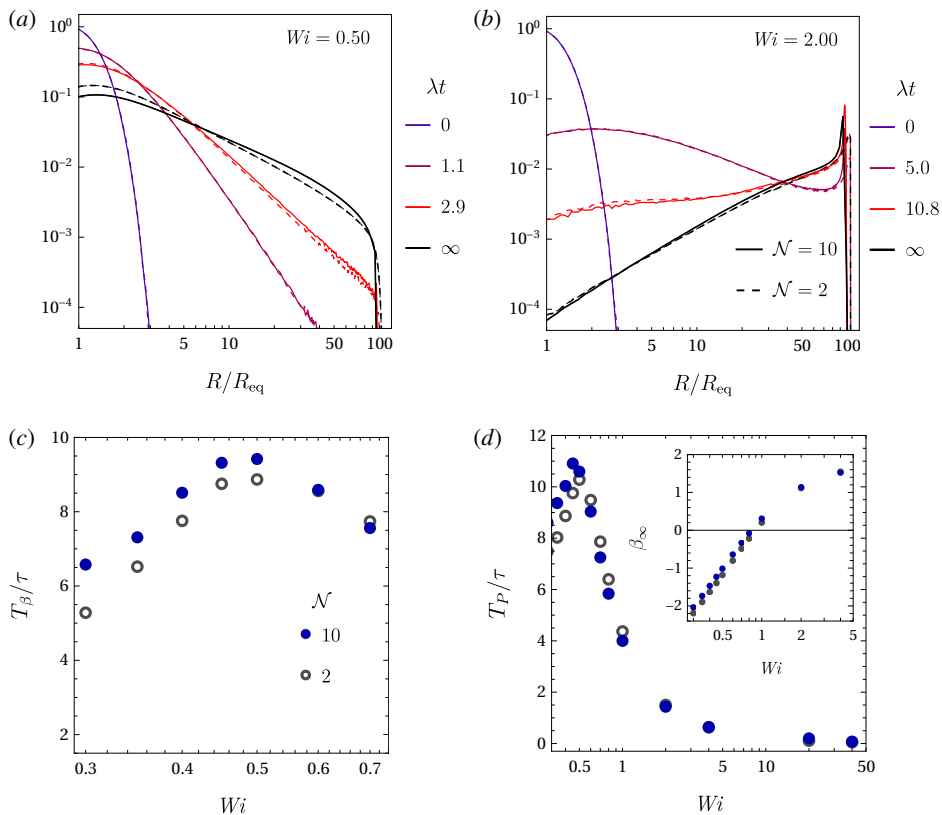

 In order to compare
the extensional dynamics of a chain and a dumbbell, \citet{jc07} proposed the
following mapping between the parameters of the two models: 
\begin{equation}\label{eq:jcmap}
    \Req = \Qeq\sqrt{\mathscr{N}-1}, \quad \Rmax = Q_m\sqrt{\mathscr{N}-1}, \quad \tau = \frac{(\mathscr{N}+1)\mathscr{N}}{6}\tau_*.
\end{equation}
This map is obtained by starting with the relation between $\Req$ and $\Qeq$, which follows from the random-walk theory for a polymer in a still fluid~\citep{deGennes79}, and then using it in an expression for the elongational viscosity of highly-stretched polymers given by \citet{wt89}. By requiring the value of the viscosity thus obtained to be independent of $\mathscr{N}$, one obtains the relation between the time-scale of the springs of an $\mathscr{N}$-bead chain, $\tau_*$, and that of an equivalent dumbbell, $\tau$ \citep{jc07}. This mapping was shown to work very well for the stationary p.d.f. of $R$ by \citet{wg10} (see the inset of figure~\ref{fig:chain}\textit{d} for a comparison of the power-law exponent corresponding to the stationary p.d.f.). Here, we check to see if it works equally well for the temporal evolution of the distribution.

Figure~\ref{fig:chain} compares the evolution and equilibration of the p.d.f. of $R$ for a dumbbell and a $\mathscr{N}=10$ chain. Panels (\textit{a}) and (\textit{b}) depict the evolution of $p(R,t)$ for a small and high $\Wi$, respectively. The results agree quite well in both cases. Figure~\ref{fig:chain}(\textit{c}) compares the equilibration time-scale of the evolving power-law, $T_\beta$, while figure~\ref{fig:chain}(\textit{d}) compares the equilibration time-scale of the entire p.d.f., $T_P$. While the results of the dumbbell and chain are similar for all $\Wi$, they are in excellent agreement at high $\Wi$. This is unsurprising, given that the mapping \eqref{eq:jcmap} was derived by equating the elongation viscosity of highly stretched chains and dumbbells~\citep{jc07}. 

\section{Concluding remarks}
\label{sec:conclusions}
Polymers in a chaotic flow field stretch and coil repeatedly, sampling a broad distribution of extensions. The basic features of the stationary distribution---a power-law tail and the coil--stretch transition---are well documented and understood in terms of the large deviations theory. In this work, we have examined the stretching dynamics in more detail to understand how the stationary distribution of extensions is attained and how polymers respond to the non-Gaussian time-correlated velocity-gradient statistics of turbulence.

We have seen that extreme strain rates are important only for stretching stiff low-$Wi$ polymers, whereas it is mild but persistent straining that extends more elastic high-$\Wi$ polymers. 
This insight has relevance for the extension of polymers in the presence of two-way coupling with the flow. The feedback forces exerted by polymers onto the flow not only reduce the mean value of the strain rate but also suppress its extreme-valued fluctuations~\citep{rll22,prasad10}. Clearly, the reduction in the mean strain rate will cause two-way coupled polymers to stretch less, on average, than one-way coupled passive polymers. In fact, at large values of $\Wi$, this effect can be strong enough to cause the mean extension to decrease with increasing $\Wi$ \citep{rpm21}. 
But what if one compensates for the reduced mean value by suitable rescaling? Would we see a separate effect of the loss of extreme events, arising from just the change in the shape of the distribution of strain-rates? 
This question was addressed in recent Eulerian--Lagrangian simulations by \citet{rll22}. They defined an effective $\Wi$ for two-way coupled polymers, using the reduced value of the Kolmogorov time scale of the modified flow, and then compared the stretching statistics of two-way and one-way coupled polymers. They found that the mean and standard deviation of the p.d.f. of extension are almost the same for one-way and two-way coupled polymers with the same effective $\Wi$. This was despite the fact that the feedback forces were strong enough to substantially modify the net dissipation rate, as well as the statistics of vorticity and strain rate. This result is understandable, indeed is to be expected, in light of our results. At small $\Wi$ where extreme strain rates are important, the feedback force and thus the modification of the flow will be weak; at large $Wi$, the extreme strain rates are no longer important for stretching and their loss does not impact the extent of stretching.
It should be noted that the simulations of \citet{rll22} were carried out at a low volume fraction;
the impact of feedback forces on polymer stretching may differ at larger volume fractions.

Examining the time evolution of the p.d.f. of extensions, we have identified two regimes: the low-$\Wi$ evolving power-law regime and the high-$\Wi$ rapid stretching regime. In both regimes, the p.d.f. equilibrates exponentially with a time-scale that peaks near the coil--stretch transition. While performing experiments or simulations at low to moderate $\Wi$, one should bear in mind that the  non-stationary p.d.f. has the form of an evolving power law, so that the mere appearance of a power-law subrange is not construed as evidence for having attained stationarity.

The turbulent carrier flow used in our simulations has a Taylor-Reynolds number of $\mathit{Re}_\lambda\approx 111$; though modest, this value is sufficient for the strain rate to exhibit strongly non-Gaussian statistics (figure~\ref{fig:persist}\textit{a}). The comparison with a Gaussian random flow has shown that the higher frequency of extreme strain rates, encountered in turbulence, has no qualitative effect on polymer stretching. So, a further intensification of the extreme strain-rates by increasing $\mathit{Re}_\lambda$ is unlikely to change the nature of stretching, although, the quantitative differences between the p.d.f. of extensions in the turbulent and Gaussian flows may be amplified.

While most of this work has considered the FENE dumbbell, we have shown that our results are valid for a bead-spring chain as well. However, we have for simplicity neglected inter-bead hydrodynamics interactions (HI) and excluded volume (EV) forces. These forces do not play a role once polymers are stretched, but they do affect the time it takes for polymers to unravel \citep{ssc04,sks07,vwrp21}. Understanding their effect on individual polymer stretching and ultimately on the turbulent dynamics of polymer solutions is an important area for future work.

One of the challenges in developing accurate models for polymers that undergo rapid stretching in turbulence is the large computational cost involved in simulating complex polymer chains in a turbulent flow. This difficulty would be greatly reduced if the DNS for the flow is replaced by a random flow field. The results of our study show that this would indeed be a useful strategy, because the stretching dynamics have been shown to be only mildly sensitive to the detailed statistics of the flow. For example, one could use a time-correlated Gaussian velocity gradient for testing the consequence of including forces like  EV and HI, and for developing coarse-grained effective dumbbell models. The latter could then be used for Lagrangian simulations in a DNS of turbulent polymer solutions.\\


\noindent \textbf{Acknowledgments.} {The authors are grateful to Samriddhi Sankar Ray for sharing his database of Lagrangian trajectories in homogeneous and isotropic turbulence.
 D.V. would like to thank the Isaac Newton Institute for Mathematical Sciences for support and hospitality during the programme
 `Mathematical aspects of turbulence: where do we stand?' when work on this paper was undertaken. J.R.P is similarly grateful for the hospitality provided by the International Centre for Theoretical Sciences (ICTS), Bangalore.
 D.V. and J.R.P. also thank the OPAL infrastructure, the Center for High-Performance Computing of Universit\'e C\^ote d’Azur, and the ICTS, Bangalore, for computational resources. Simulations were also performed on the IIT Bombay workstation \textit{Aragorn}.}\\

 \noindent \textbf{Funding.} {This work was supported by IRCC, IIT Bombay (J.R.P., seed grant); DST-SERB (J.R.P., grant number SRG/2021/001185);
 F\'ed\'eration de Recherche Wolfgang Doeblin (J.R.P.);
 the Agence Nationale de la Recherche (D.V., grant numbers ANR-21-CE30-0040-01, ANR-15-IDEX-01);
 EPSRC (D.V., grant number EP/R014604/1); the Simons Foundation (D.V.); and the Indo--French Centre for Applied Mathematics IFCAM (J.R.P. and D.V.).}\\

\noindent \textbf{Declaration of interests.} {The authors report no conflict of interest.}\\


\noindent \textbf{Author ORCID} {J.~R.~Picardo, https://orcid.org/0000-0002-9227-5516; 
 E.~L.~C.~VI~M. Plan, https://orcid.org/0000-0003-2268-424X;
 D.~Vincenzi, https://orcid.org/0000-0003-3332-3802}

\bibliographystyle{jfm}
\bibliography{polymers}

\begin{thebibliography}{76}
\expandafter\ifx\csname natexlab\endcsname\relax\def\natexlab#1{#1}\fi
\def\au#1{#1} \def\ed#1{#1} \def\yr#1{#1}\def\at#1{#1}\def\jt#1{\textit{#1}}
  \def\bt#1{#1}\def\bvol#1{\textbf{#1}} \def\vol#1{#1} \def\pg#1{#1}
  \def\publ#1{#1}\def\arxiv#1{#1}\def\org#1{#1}\def\st#1{\textit{#1}}

\bibitem[Anand {\em et~al.\/}(2020)Anand, Ray \& Subramanian]{ars20}
{\normalfont \au{Anand, P.}, \au{Ray, S.~S.} \& \au{Subramanian, G.}} \yr{2020}
   \at{Orientation dynamics of sedimenting anisotropic particles in
  turbulence}.  \jt{Phys. Rev. Lett.}  \bvol{125},  \pg{034501}.

\bibitem[Bagheri {\em et~al.\/}(2012)Bagheri, Mitra, Perlekar \&
  Brandt]{bmpb12}
{\normalfont \au{Bagheri, F.}, \au{Mitra, D.}, \au{Perlekar, P.} \& \au{Brandt,
  L.}} \yr{2012}  \at{Statistics of polymer extensions in turbulent channel
  flow}.  \jt{Phys. Rev. E}  \bvol{86},  \pg{056314}.

\bibitem[Balkovsky {\em et~al.\/}(2000)Balkovsky, Fouxon \& Lebedev]{bfl00}
{\normalfont \au{Balkovsky, E.}, \au{Fouxon, A.} \& \au{Lebedev, V.}} \yr{2000}
   \at{Turbulent dynamics of polymer solutions}.  \jt{Phys. Rev. Lett.}
  \bvol{84},  \pg{4765--4768}.

\bibitem[Balkovsky {\em et~al.\/}(2001)Balkovsky, Fouxon \& Lebedev]{bfl01}
{\normalfont \au{Balkovsky, E.}, \au{Fouxon, A.} \& \au{Lebedev, V.}} \yr{2001}
   \at{Turbulence of polymer solutions}.  \jt{Phys. Rev. E}  \bvol{64},
  \pg{056301}.

\bibitem[Bec {\em et~al.\/}(2006)Bec, Biferale, Boffetta, Cencini, Musacchio \&
  Toschi]{bbbcmt06}
{\normalfont \au{Bec, J.}, \au{Biferale, L.}, \au{Boffetta, G.}, \au{Cencini,
  M.}, \au{Musacchio, S.} \& \au{Toschi, F.}} \yr{2006}  \at{{Lyapunov}
  exponents of heavy particles in turbulence}.  \jt{Phys. Fluids}  \bvol{18},
  \pg{091702}.

\bibitem[Benzi \& Ching(2018)]{bc18}
{\normalfont \au{Benzi, R.} \& \au{Ching, E.~S.~C.}} \yr{2018}  \at{Polymers in
  fluid flows}.  \jt{Annu. Rev. Condens. Matter Phys.}  \bvol{9},
  \pg{163--181}.

\bibitem[Bhatnagar {\em et~al.\/}(2016)Bhatnagar, Gupta, Mitra, Pandit \&
  Perlekar]{Akshay}
{\normalfont \au{Bhatnagar, A.}, \au{Gupta, A.}, \au{Mitra, D.}, \au{Pandit,
  R.} \& \au{Perlekar, P.}} \yr{2016}  \at{How long do particles spend in
  vortical regions in turbulent flows?}  \jt{Phys. Rev. E}  \bvol{94},
  \pg{053119}.

\bibitem[Biferale {\em et~al.\/}(2014)Biferale, Meneveau \& Verzicco]{bmv14}
{\normalfont \au{Biferale, L.}, \au{Meneveau, C.} \& \au{Verzicco, R.}}
  \yr{2014}  \at{Deformation statistics of sub-{K}olmogorov-scale ellipsoidal
  neutrally buoyant drops in isotropic turbulence}.  \jt{J. Fluid Mech.}
  \bvol{754},  \pg{184--207}.

\bibitem[Bird {\em et~al.\/}(1987)Bird, Curtiss, Armstrong \& Hassager]{bird}
{\normalfont \au{Bird, R.~B.}, \au{Curtiss, C.~F.}, \au{Armstrong, R.~C.} \&
  \au{Hassager, O.}} \yr{1987} {\em Dynamics of Polymeric Liquids\/}, ,
  \vol{vol.~2}.  \publ{Wiley}.

\bibitem[Boffetta {\em et~al.\/}(2003)Boffetta, Celani \& Musacchio]{bcm03}
{\normalfont \au{Boffetta, G.}, \au{Celani, A.} \& \au{Musacchio, S.}}
  \yr{2003}  \at{Two-dimensional turbulence of dilute polymer solutions}.
  \jt{Phys. Rev. Lett.}  \bvol{91},  \pg{034501}.

\bibitem[Bray {\em et~al.\/}(2013)Bray, Majumdar \& Schehr]{Bray}
{\normalfont \au{Bray, Alan~J.}, \au{Majumdar, Satya~N.} \& \au{Schehr,
  Grégory}} \yr{2013}  \at{Persistence and first-passage properties in
  nonequilibrium systems}.  \jt{Advances in Physics}  \bvol{62}~(3),
  \pg{225--361}.

\bibitem[Brunk {\em et~al.\/}(1997)Brunk, Koch \& Lion]{bkl97}
{\normalfont \au{Brunk, B.~K.}, \au{Koch, D.~L.} \& \au{Lion, L.~W.}} \yr{1997}
   \at{Hydrodynamic pair diffusion in isotropic random velocity fields with
  application to turbulent coagulation}.  \jt{Phys. Fluids}  \bvol{9},
  \pg{2670--2691}.

\bibitem[Cecconi {\em et~al.\/}(2010)Cecconi, Cencini \& Vulpiani]{cc2010}
{\normalfont \au{Cecconi, F.}, \au{Cencini, M.} \& \au{Vulpiani, A.}} \yr{2010}
  {\em Chaos: from Simple Models to Complex Systems\/}.  \publ{Singapore: World
  Scientific}.

\bibitem[Celani {\em et~al.\/}(2006)Celani, Puliafito \& Vincenzi]{cpv06}
{\normalfont \au{Celani, A.}, \au{Puliafito, A.} \& \au{Vincenzi, D.}}
  \yr{2006}  \at{Dynamical slowdown of polymers in laminar and random flows}.
  \jt{Phys. Rev. Lett.}  \bvol{97},  \pg{118301}.

\bibitem[Chertkov(2000)]{c00}
{\normalfont \au{Chertkov, M.}} \yr{2000}  \at{Polymer stretching by
  turbulence}.  \jt{Phys. Rev. Lett.}  \bvol{84},  \pg{4761--4764}.

\bibitem[Chevillard \& Meneveau(2013)]{cm13}
{\normalfont \au{Chevillard, L.} \& \au{Meneveau, C.}} \yr{2013}
  \at{Orientation dynamics of small, triaxial–ellipsoidal particles in
  isotropic turbulence}.  \jt{J. Fluid Mech.}  \bvol{737},  \pg{571--596}.

\bibitem[Childress \& Gilbert(1995)]{cg95}
{\normalfont \au{Childress, S.} \& \au{Gilbert, A.~D.}} \yr{1995} {\em Stretch,
  Twist, Fold: The Fast Dynamo\/}.  \publ{Berlin: Springer}.

\bibitem[Dieci {\em et~al.\/}(1997)Dieci, Russell \& Vleck]{drv97}
{\normalfont \au{Dieci, L.}, \au{Russell, R.~D.} \& \au{Vleck, E. S.~Van}}
  \yr{1997}  \at{On the computation of {L}yapunov exponents for continuous
  dynamical systems}.  \jt{SIAM J. Numer. Anal.}  \bvol{34},  \pg{402--423}.

\bibitem[Eckhardt {\em et~al.\/}(2002)Eckhardt, Kronj{\"a}ger \&
  Schumacher]{eks02}
{\normalfont \au{Eckhardt, B.}, \au{Kronj{\"a}ger, J.} \& \au{Schumacher, J.}}
  \yr{2002}  \at{Stretching of polymers in a turbulent environment}.
  \jt{Comput. Phys. Commun.}  \bvol{147},  \pg{538--543}.

\bibitem[Falkovich {\em et~al.\/}(2001)Falkovich, Gaw\c{e}dki \&
  Vergassola]{fgv01}
{\normalfont \au{Falkovich, G.}, \au{Gaw\c{e}dki, K.} \& \au{Vergassola, M.}}
  \yr{2001}  \at{Particles and fields in fluid turbulence}.  \jt{Rev. Mod.
  Phys.}  \bvol{73},  \pg{913--975}.

\bibitem[Frisch(1995)]{f95}
{\normalfont \au{Frisch, U.}} \yr{1995} {\em Turbulence: The Legacy of A. N.
  Kolmogorov\/}.  \publ{Cambridge, England: Cambridge University Press}.

\bibitem[de~Gennes(1979)]{deGennes79}
{\normalfont \au{de~Gennes, P.-G.}} \yr{1979} {\em Scaling Concepts in Polymer
  Physics\/}.  \publ{Ithaca, NY: Cornell University Press}.

\bibitem[Gerashchenko {\em et~al.\/}(2005)Gerashchenko, Chevallard \&
  Steinberg]{gcs05}
{\normalfont \au{Gerashchenko, S}, \au{Chevallard, C.} \& \au{Steinberg, V.}}
  \yr{2005}  \at{Single-polymer dynamics: Coil-stretch transition in a random
  flow}.  \jt{Europhys. Lett.}  \bvol{71},  \pg{221--227}.

\bibitem[Gerashchenko \& Steinberg(2008)]{gs08}
{\normalfont \au{Gerashchenko, S} \& \au{Steinberg, V.}} \yr{2008}
  \at{Critical slowing down in polymer dynamics near the coil-stretch
  transition in elongation flow}.  \jt{Phys. Rev. E}  \bvol{78},
  \pg{040801(R)}.

\bibitem[Graham(2014)]{g14}
{\normalfont \au{Graham, M.~D.}} \yr{2014}  \at{Drag reduction and the dynamics
  of turbulence in simple and complex fluids}.  \jt{Phys. Fluids}  \bvol{26},
  \pg{101301}.

\bibitem[Graham(2018)]{g18}
{\normalfont \au{Graham, M.~D.}} \yr{2018} {\em Microhydrodynamics, Brownian
  Motion, and Complex Fluids\/}.  \publ{Cambridge, UK: Cambridge University
  Press}.

\bibitem[Groisman \& Steinberg(2001)]{gsprl01}
{\normalfont \au{Groisman, A.} \& \au{Steinberg, V.}} \yr{2001}  \at{Stretching
  of polymers in a random three-dimensional flow}.  \jt{Phys. Rev. Lett.}
  \bvol{86},  \pg{934--937}.

\bibitem[Gupta {\em et~al.\/}(2015)Gupta, Perlekar \& Pandit]{gpp15}
{\normalfont \au{Gupta, A.}, \au{Perlekar, P.} \& \au{Pandit, R.}} \yr{2015}
  \at{Two-dimensional homogeneous isotropic fluid turbulence with polymer
  additives}.  \jt{Phys. Rev. E}  \bvol{91},  \pg{033013}.

\bibitem[Gupta {\em et~al.\/}(2004)Gupta, Sureshkumar \& Khomami]{gsk04}
{\normalfont \au{Gupta, V.~K.}, \au{Sureshkumar, R.} \& \au{Khomami, B.}}
  \yr{2004}  \at{Polymer chain dynamics in {N}ewtonian and viscoelastic
  turbulent channel flows}.  \jt{Phys. Fluids}  \bvol{16},  \pg{1546}.

\bibitem[Gustavsson {\em et~al.\/}(2014)Gustavsson, Einarsson \& Mehlig]{gem14}
{\normalfont \au{Gustavsson, K.}, \au{Einarsson, J.} \& \au{Mehlig, B.}}
  \yr{2014}  \at{Tumbling of small axisymmetric particles in random and
  turbulent flows}.  \jt{Phys. Rev. Lett.}  \bvol{112},  \pg{014501}.

\bibitem[Ilg {\em et~al.\/}(2002)Ilg, Angelis, Karlin, Casciola \&
  Succi]{idakcs02}
{\normalfont \au{Ilg, P.}, \au{Angelis, E.~De}, \au{Karlin, I.~V.},
  \au{Casciola, C.~M.} \& \au{Succi, S.}} \yr{2002}  \at{Polymer dynamics in
  wall turbulent flow}.  \jt{Europhys. Lett.}  \bvol{58},  \pg{616}.

\bibitem[James \& Ray(2017)]{jr17}
{\normalfont \au{James, M.} \& \au{Ray, S.~S.}} \yr{2017}  \at{Enhanced droplet
  collision rates and impact velocities in turbulent flows: The effect of
  poly-dispersity and transient phases}.  \jt{Sci. Reports}  \bvol{7},
  \pg{12231}.

\bibitem[Jin \& Collins(2007)]{jc07}
{\normalfont \au{Jin, S.} \& \au{Collins, L.~R.}} \yr{2007}  \at{Dynamics of
  dissolved polymer chains in isotropic turbulence}.  \jt{New J. Phys.}
  \bvol{9},  \pg{360}.

\bibitem[Kadoch {\em et~al.\/}(2011)Kadoch, del Castillo-Negrete, Bos \&
  Schneider]{Bos}
{\normalfont \au{Kadoch, B.}, \au{del Castillo-Negrete, D.}, \au{Bos, W. J.~T.}
  \& \au{Schneider, K.}} \yr{2011}  \at{Lagrangian statistics and flow topology
  in forced two-dimensional turbulence}.  \jt{Phys. Rev. E}  \bvol{83},
  \pg{036314}.

\bibitem[Liu \& Steinberg(2010)]{ls10}
{\normalfont \au{Liu, Y.} \& \au{Steinberg, V.}} \yr{2010}  \at{Stretching of
  polymer in a random flow: Effect of a shear rate}.  \jt{Europhys. Lett.}
  \bvol{90},  \pg{44005}.

\bibitem[Liu \& Steinberg(2014)]{ls14}
{\normalfont \au{Liu, Y.} \& \au{Steinberg, V.}} \yr{2014}  \at{Single polymer
  dynamics in a random flow}.  \jt{Macromol. Symp.}  \bvol{337},  \pg{34--43}.

\bibitem[Lumley(1973)]{l73}
{\normalfont \au{Lumley, J.~L.}} \yr{1973}  \at{Drag reduction in turbulent
  flow by polymer additives}.  \jt{J. Polymer Sci. Macromol. Rev.}  \bvol{7},
  \pg{263--290}.

\bibitem[Majumdar(1999)]{Satya-Review}
{\normalfont \au{Majumdar, S.~N.}} \yr{1999}  \at{Persistence in nonequilibrium
  systems}.  \jt{Current Science}  \bvol{77}~(3),  \pg{370--375}.

\bibitem[{Martins Afonso} \& Vincenzi(2005)]{mav05}
{\normalfont \au{{Martins Afonso}, M.} \& \au{Vincenzi, D.}} \yr{2005}
  \at{Nonlinear elastic polymers in random flow}.  \jt{J. Fluid Mech.}
  \bvol{540},  \pg{99--108}.

\bibitem[Massah {\em et~al.\/}(1993)Massah, Kontomaris, Schowalter \&
  Hanratty]{mksh93}
{\normalfont \au{Massah, H.}, \au{Kontomaris, K.}, \au{Schowalter, W.~R.} \&
  \au{Hanratty, T.~J.}} \yr{1993}  \at{The configurations of a {FENE}
  bead‐spring chain in transient rheological flows and in a turbulent flow}.
  \jt{Phys. Fluids A}  \bvol{5},  \pg{881}.

\bibitem[Mosler \& Shaqfeh(1997)]{ms97}
{\normalfont \au{Mosler, A.~B.} \& \au{Shaqfeh, E.~S.~G.}} \yr{1997}  \at{The
  conformation change of model polymers in stochastic flow fields: Flow through
  fixed beds}.  \jt{Phys. Fluids}  \bvol{9},  \pg{1222}.

\bibitem[Nguyen \& Kausch(1999)]{nk99}
{\normalfont \au{Nguyen, T.~Q.} \& \au{Kausch, H.-H.}}, ed. \yr{1999} {\em
  Flexible Polymer Chain Dynamics in Elongational Flow\/}.  \publ{Berlin
  Heidelberg: Springer}.

\bibitem[\"Ottinger(1996)]{o96}
{\normalfont \au{\"Ottinger, H.~C.}} \yr{1996} {\em Stochastic Processes in
  Polymeric Fluids\/}.  \publ{Berlin: Springer}.

\bibitem[Perkins {\em et~al.\/}(1997)Perkins, Smith \& Chu]{psc97}
{\normalfont \au{Perkins, T.}, \au{Smith, D.~E.} \& \au{Chu, S.}} \yr{1997}
  \at{Single polymer dynamics in an elongational flow}.  \jt{Science}
  \bvol{276},  \pg{2016--2021}.

\bibitem[Perlekar {\em et~al.\/}(2010)Perlekar, Mitra \& Pandit]{prasad10}
{\normalfont \au{Perlekar, P.}, \au{Mitra, D.} \& \au{Pandit, R.}} \yr{2010}
  \at{Direct numerical simulations of statistically steady, homogeneous,
  isotropic fluid turbulence with polymer additives}.  \jt{Phys. Rev. E}
  \bvol{82},  \pg{066313}.

\bibitem[Perlekar {\em et~al.\/}(2011)Perlekar, Ray, Mitra \& Pandit]{Prasad}
{\normalfont \au{Perlekar, P.}, \au{Ray, S.~S.}, \au{Mitra, D.} \& \au{Pandit,
  R.}} \yr{2011}  \at{Persistence problem in two-dimensional fluid turbulence}.
   \jt{Phys. Rev. Lett.}  \bvol{106},  \pg{054501}.

\bibitem[Peters \& Schumacher(2007)]{ps07}
{\normalfont \au{Peters, T.} \& \au{Schumacher, J.}} \yr{2007}  \at{Two-way
  coupling of finitely extensible nonlinear elastic dumbbells with a turbulent
  shear flow}.  \jt{Phys. Fluids}  \bvol{19},  \pg{065109}.

\bibitem[Petzold(1983)]{lsodaref}
{\normalfont \au{Petzold, L.}} \yr{1983}  \at{Automatic selection of methods
  for solving stiff and nonstiff systems of ordinary differential equations}.
  \jt{SIAM Journal on Scientific and Statistical Computing}  \bvol{4}~(1),
  \pg{136--148}.

\bibitem[Plan {\em et~al.\/}(2016)Plan, Ali \& Vincenzi]{pav16}
{\normalfont \au{Plan, E. L. C. VI~M.}, \au{Ali, A.} \& \au{Vincenzi, D.}}
  \yr{2016}  \at{Bead-rod-spring models in random flows}.  \jt{Phys. Rev. E}
  \bvol{94},  \pg{020501(R)}.

\bibitem[Puliafito \& Turitsyn(2005)]{pt05}
{\normalfont \au{Puliafito, A.} \& \au{Turitsyn, K.}} \yr{2005}  \at{Numerical
  study of polymer tumbling in linear shear flows}.  \jt{Physica D}
  \bvol{211},  \pg{9--22}.

\bibitem[Pumir \& Wilkinson(2011)]{pw11}
{\normalfont \au{Pumir, A.} \& \au{Wilkinson, M.}} \yr{2011}  \at{Orientation
  statistics of small particles in turbulence}.  \jt{New J. Phys.}  \bvol{13},
  \pg{093030}.

\bibitem[ur~Rehman {\em et~al.\/}(2022)ur~Rehman, Lee \& Lee]{rll22}
{\normalfont \au{ur~Rehman, S.}, \au{Lee, J.} \& \au{Lee, C.}} \yr{2022}
  \at{Effect of {Weissenberg} number on polymer-laden turbulence}.  \jt{Phys.
  Rev. Fluids}  \bvol{7},  \pg{064303}.

\bibitem[Rosti {\em et~al.\/}(2021)Rosti, Perlekar \& Mitra]{rpm21}
{\normalfont \au{Rosti, M.~E.}, \au{Perlekar, P.} \& \au{Mitra, D.}} \yr{2021}
  Large is different: non-monotonic behaviour of elastic range scaling in
  polymeric turbulence at large {R}eynolds and {D}eborah numbers.
  \texttt{arXiv:2111.11224}.

\bibitem[Schroeder(2018)]{s18}
{\normalfont \au{Schroeder, C.~M.}} \yr{2018}  \at{Single polymer dynamics for
  molecular rheology}.  \jt{J. Rheol.}  \bvol{62},  \pg{371--403}.

\bibitem[Schroeder {\em et~al.\/}(2003)Schroeder, Babcock, Shaqfeh \&
  Chu]{sbsc03}
{\normalfont \au{Schroeder, C.~M.}, \au{Babcock, H.~P.}, \au{Shaqfeh, E. S.~G.}
  \& \au{Chu, S.}} \yr{2003}  \at{Observation of polymer conformation
  hysteresis in extensional flow}.  \jt{Science}  \bvol{301},  \pg{1515--1519}.

\bibitem[Schroeder {\em et~al.\/}(2004)Schroeder, Shaqfeh \& Chu]{ssc04}
{\normalfont \au{Schroeder, C.~M.}, \au{Shaqfeh, E.~S.~G.} \& \au{Chu, S.}}
  \yr{2004}  \at{Effect of hydrodynamic interactions on dna dynamics in
  extensional flow: Simulation and single molecule experiment}.
  \jt{Macromolecules}  \bvol{37},  \pg{9242--9256}.

\bibitem[Serafini {\em et~al.\/}(2022)Serafini, Battista, Gualtieri \&
  Casciola]{sbgc22}
{\normalfont \au{Serafini, F.}, \au{Battista, F.}, \au{Gualtieri, P.} \&
  \au{Casciola, C.~M.}} \yr{2022}  \at{Drag reduction in turbulent wall-bounded
  flows of realistic polymer solutions}.  \jt{Phys. Rev. Lett.}  \bvol{129},
  \pg{104502}.

\bibitem[Shaqfeh \& Koch(1992)]{sk92}
{\normalfont \au{Shaqfeh, E. S.~G.} \& \au{Koch, D.~L.}} \yr{1992}  \at{Polymer
  stretch in dilute fixed beds of fibres or spheres}.  \jt{J. Fluid Mech.}
  \bvol{244},  \pg{17--54}.

\bibitem[Sim {\em et~al.\/}(2007)Sim, Khomami \& Sureshkumar]{sks07}
{\normalfont \au{Sim, H.~G.}, \au{Khomami, B.} \& \au{Sureshkumar, R.}}
  \yr{2007}  \at{Flow–induced chain scission in dilute polymer solutions:
  Algorithm development and results for scission dynamics in elongational
  flow}.  \jt{J. Rheol.}  \bvol{51}~(6),  \pg{1223--1251}.

\bibitem[Singh {\em et~al.\/}(2022)Singh, Picardo \& Ray]{settling}
{\normalfont \au{Singh, R.~K.}, \au{Picardo, J.~R.} \& \au{Ray, S.~S.}}
  \yr{2022}  \at{Sedimenting elastic filaments in turbulent flows}.  \jt{Phys.
  Rev. Fluids}  \bvol{7},  \pg{084502}.

\bibitem[Soares(2020)]{s20}
{\normalfont \au{Soares, E.~J.}} \yr{2020}  \at{Review of mechanical
  degradation and de-aggregation of drag reducing polymers in turbulent flows}.
   \jt{J. Non-Newtonian Fluid Mech.}  \bvol{276},  \pg{104225}.

\bibitem[Steinberg(2021)]{steinberg21}
{\normalfont \au{Steinberg, V.}} \yr{2021}  \at{Elastic turbulence: An
  experimental view on inertialess random flow}.  \jt{Annu. Rev. Fluid Mech.}
  \bvol{53},  \pg{27--58}.

\bibitem[Stone \& Graham(2003)]{sg03}
{\normalfont \au{Stone, P.~A.} \& \au{Graham, M.~D.}} \yr{2003}  \at{Polymer
  dynamics in a model of the turbulent buffer layer}.  \jt{Phys. Fluids}
  \bvol{15},  \pg{1247--1256.}

\bibitem[Terrapon {\em et~al.\/}(2004)Terrapon, Dubief, Moin, Shaqfeh \&
  Lele]{tdmsl04}
{\normalfont \au{Terrapon, V.~E.}, \au{Dubief, Y.}, \au{Moin, P.}, \au{Shaqfeh,
  E. S.~G.} \& \au{Lele, S.~K.}} \yr{2004}  \at{Simulated polymer stretch in a
  turbulent flow using {B}rownian dynamics}.  \jt{J. Fluid Mech.}  \bvol{504},
  \pg{61}.

\bibitem[Vanneste(2010)]{v10}
{\normalfont \au{Vanneste, J.}} \yr{2010}  \at{Estimating generalized
  {L}yapunov exponents for products of random matrices}.  \jt{Phys. Rev. E}
  \bvol{81},  \pg{036701}.

\bibitem[Vincenzi {\em et~al.\/}(2021)Vincenzi, Watanabe, Ray \&
  Picardo]{vwrp21}
{\normalfont \au{Vincenzi, D.}, \au{Watanabe, T.}, \au{Ray, S.~S.} \&
  \au{Picardo, J.}} \yr{2021}  \at{Polymer scission in turbulent flows}.
  \jt{J. Fluid Mech.}  \bvol{912},  \pg{A18}.

\bibitem[Watanabe \& Gotoh(2010)]{wg10}
{\normalfont \au{Watanabe, T.} \& \au{Gotoh, T.}} \yr{2010}  \at{Coil-stretch
  transition in an ensemble of polymers in isotropic turbulence}.  \jt{Phys.
  Rev. E}  \bvol{81},  \pg{066301}.

\bibitem[Watanabe \& Gotoh(2013)]{wg13}
{\normalfont \au{Watanabe, T.} \& \au{Gotoh, T.}} \yr{2013}  \at{Hybrid
  {E}ulerian--{L}agrangian simulations for polymer–turbulence interactions}.
  \jt{J. Fluid Mech.}  \bvol{717},  \pg{535–575}.

\bibitem[Wiest \& Tanner(1989)]{wt89}
{\normalfont \au{Wiest, J.~M.} \& \au{Tanner, R.~I.}} \yr{1989}  \at{Rheology
  of bead‐nonlinear spring chain macromolecules}.  \jt{J. Rheol.}
  \bvol{33}~(2),  \pg{281--316}.

\bibitem[{Wolfram Research}(2019)]{ndsolve}
{\normalfont \au{{Wolfram Research}}} \yr{2019} {NDSolve}.
  \url{https://reference.wolfram.com/language/ref/NDSolve.html}.

\bibitem[Xi(2019)]{x19}
{\normalfont \au{Xi, L.}} \yr{2019}  \at{Turbulent drag reduction by polymer
  additives: Fundamentals and recent advances}.  \jt{Phys. Fluids}  \bvol{31},
  \pg{121302}.

\bibitem[Yeung(2001)]{y01}
{\normalfont \au{Yeung, P.~K.}} \yr{2001}  \at{Lagrangian characteristics of
  turbulence and scalar transport in direct numerical simulations}.  \jt{J.
  Fluid Mech.}  \bvol{427},  \pg{241--274}.

\bibitem[Young(1999)]{y99}
{\normalfont \au{Young, W.~R.}} \yr{1999} Stirring and mixing.  \bt{In {\em
  1999 Summer Program in Geophysical Fluid Dynamics\/} (ed. \ed{J.-L.
  Thiffeault \& C.~Pasquero})}. Woods Hole Oceanographic Institution, Woods
  Hole, MA.

\bibitem[Young(2009)]{y09}
{\normalfont \au{Young, W.~R.}} \yr{2009} The passive scalar problem.
  {I}nstitut Henri Poincar\'e, Paris, \url{https://mhd.ens.fr/IHP09/Young/}.

\bibitem[Zel'dovich {\em et~al.\/}(1984)Zel'dovich, Ruzmaikin, Molchanov \&
  Sokoloff]{zrms84}
{\normalfont \au{Zel'dovich, Ya.~B.}, \au{Ruzmaikin, A.~A.}, \au{Molchanov,
  S.~A.} \& \au{Sokoloff, D.~D.}} \yr{1984}  \at{Kinematic dynamo problem in a
  linear velocity field}.  \jt{J. Fluid Mech.}  \bvol{144},  \pg{1--11}.

\bibitem[Zhou \& Akhavan(2003)]{za03}
{\normalfont \au{Zhou, Q.} \& \au{Akhavan, R.}} \yr{2003}  \at{A comparison of
  {FENE} and {FENE-P} dumbbell and chain models in turbulent flow}.  \jt{J.
  Non-Newtonian Fluid Mech.}  \bvol{109},  \pg{115}.

\end{thebibliography}

\end{document}